\documentclass[smallextended]{svjour3_adapt} 
\usepackage[backend=biber, style=numeric, citestyle=numeric-comp]{biblatex}
\usepackage{a4wide, graphicx, url, fancyhdr, amssymb, amsmath, mathrsfs, float, hyperref, amsfonts,bm,ulem}
\usepackage{color,soul}
\usepackage{lmodern,multicol}
\usepackage[utf8]{inputenc}
\usepackage[T1]{fontenc}
\usepackage{subcaption}
\usepackage{blkarray}
\usepackage{float}
\usepackage{bbm}
\usepackage{multirow}
\usepackage{multicol}
\usepackage{booktabs}
\usepackage{pgfplots}
\usepackage{pgfplotstable}
\usepackage{setspace}
\usepackage{caption}
\usepackage{chngcntr}

\usepackage{arydshln}
\usepackage{rotating}

\DeclareMathOperator*{\argmax}{\arg\!\max}

\begin{document}

\title{Binary disease prediction using tail quantiles of the distribution of continuous biomarkers
}
\author{Michiel H.J. ~Paus         \and
        Edwin R. van den ~Heuvel   \and
        Marc J.M. ~Meddens
}


\institute{Michiel H.J. ~Paus \at
              Department of Mathematics and Computer Science, Eindhoven University of Technology, the Netherlands \\
              \email{m.h.j.paus@outlook.com}           
           \and
            Edwin R. van den ~Heuvel \at
            Department of Mathematics and Computer Science, Eindhoven University of Technology, the Netherlands \\
            \email{e.r.v.d.heuvel@tue.nl}          
            \and
            Marc J.M. ~Meddens \at
            Brainscan BV Deventer, the Netherlands \\
            \email{marcmeddens@brainscanbv.com}
}

\date{\today}


\maketitle

\begin{abstract}
In the analysis of binary disease classification, single biomarkers might not have significant discriminating power and multiple biomarkers from a large set of biomarkers should be selected. Many different approaches exist, but they merely work well for mean differences in biomarkers between cases and controls. Biological processes are however much more heterogeneous, and differences between cases and controls could also occur in other distributional characteristics (e.g. variances, skewness). Many machine learning techniques are better capable of utilizing these higher order distributional differences, sometimes at cost of explainability.

In this study we propose quantile based prediction (QBP), a binary classification method that is based on the selection of multiple continuous biomarkers. It can be considered a hybrid technique, with the flexibility of a machine learning algorithm and the ability to select relevant features like classical statistical techniques. QBP generates a single score using the tails of the biomarker distributions for cases and controls. This single score can then be evaluated by receiver operating characteristic (ROC) analysis to investigate its predictive power. 

The performance of QBP is compared to supervised learning methods using extensive simulation studies, and two case studies: major depression disorder (MDD) and trisomy. Simultaneously, the classification performance of the existing techniques in relation to each other is assessed. The key strengths of QBP are the opportunity to select relevant biomarkers and the outstanding classification performance in the case biomarkers predominantly show variance differences between cases and controls, as demonstrated in the simulation study. When only shifts in means were present in the biomarkers, QBP obtained an inferior performance. Lastly, QBP proved to be unbiased in case of absence of disease relevant biomarkers and outperformed the other methods on the MDD case study. 

More research is needed to further optimize QBP, since it has several opportunities to improve its performance. Here we wanted to introduce the principle of QBP and show its potential.


\end{abstract}


\newpage
\normalsize
\section{Introduction}
Biomarker research has increased fastly due to the development of new molecular biotechnologies \cite{pepe2008pivotal}. A biomarker is defined as '\textit{any substance, structure, or process that can be measured in the body or its products and influence or predict the incidence of outcome or disease}' \cite{world2001biomarkers}. Biomarkers are developed for many different purposes: classification and prediction of diseases, as surrogate outcomes in clinical trials, as measures of toxic or preventive exposures, or as a guide to individual treatment choice \cite{halaris2013inflammation}. \\

For the classification and prediction of diseases, single biomarkers do often not have sufficient discriminating power to separate cases from controls \cite{calfee2011use, hsu2014biomarker, jentsch2015biomarker}. When analyzing multiple biomarkers simultaneously, models might become harder to interpret, but could also face the problem of high dimensionality with respect to the available number of observations. 

Firstly, to enhance the transparency of these classification or prediction models with numerous biomarkers, insight in the selected features and its importance is crucial. Whereas classical statistical techniques hold the possibility to perform in-depth inference on present relations, many machine learning techniques do not allow for a similar degree of interpretability. Secondly, when the number of biomarkers $p$ exceeds the number of observations $n$ ($p>n$ or $p>>n$), it is key to reduce the dimensionality of the data and to select a sparse set of biomarkers with high discriminant power that can be used to produce reliable predictions. 

Binary classification methods that reduce the dimensionality of the input variables can be categorized based on the relations between original input variables and new input variables \cite{ma2008penalized}. (i) Dimension reduction methods that construct new input variables using linear combinations of $all$ input variables (i.e. partial least squares (PLS) and principal component analysis (PCA). (ii) Feature selection methods, which select a subset of the original input variables. Examples include likelihood functions for parametric models, such as penalized logistic regression (PLR) and linear discriminant analysis (LDA) by optimal scoring. (iii) Hybrid methods using (i) and (ii). These traditional methods, focus mainly on mean differences of the biomarker distributions between cases and controls. However, differences may occur somewhere else, since a disease may affect the variation, skewness and kurtosis of the biomarker distribution \cite{just2014improving}. 

Over time, a wide scala of classification tree based techniques is developed, from individual trees (CART) to an ensemble of individual trees with various modifications such as different sampling strategies like bootstrapping (Random Forest) or boosting (AdaBoost, XGBoost). Other machine learning techniques for classification include support vector machines (SVM) and the k-nearest neighbors (kNN) algorithm that does not require a model to be fit \cite{friedman2001elements}.\\

In this paper we introduce a new approach for binary classification that takes advantage of the tail differences of the biomarker distributions between cases and controls. The performance of this new method is compared with various traditional binary classification methods and machine learning techniques using simulation studies and two case studies. Logistic regression is applied with and without penalization. The selected penalty functions are the lasso \cite{tibshirani1996regression}, elastic net \cite{zou2005regularization} and the ridge \cite{hoerl1970ridge}. Alternatively, to address multicollinearity among the predictors, principal component logistic regression (PCLR) is included in the analysis \cite{aguilera2006using}. Next to these LR based methods, also LDA and PLS with LDA (abbreviated as PLS-LDA) was used \cite{marigheto1998comparison}. The considered machine learning techniques include SVM, kNN, random forest (RF) and extreme gradient boosting (XGBoost).

The first case study describes data on patients with major depressive disorder (MDD), which is a disease with a lifetime prevalence of around 15\%. It is a major cause of disability in the Western world \cite{bromet2011cross,sobocki2006cost} and the prediction of MDD with biomarkers can help physicians diagnose MDD better. The second case study, is on an ongoing Dutch population study on the prevalence of trisomy 13, 18 and 21, containing 4894 observations.

In this paper, the receiver operating characteristic (ROC) curve approach is used to derive the classification performance of cases and controls. In specific, we measure the area under the ROC curve (AUC). The AUC is a variant of the concordance ($c$) statistic for binary outcomes, that indicates the discriminative ability of a generalized linear model \cite{steyerberg2010assessing}. Advantages of this non-parametric statistic are that it does not depend on a decision threshold and gives an indication of how well the negative and positive classes are separated \cite{bradley1997use}.

To assess the predictive performance of all methods in terms of AUC, we use different cross-validation strategies. For the simulation scenarios we apply k-fold cross-validation (CV) on the training dataset to determine the set of tunable parameters with the highest average AUC over all k folds. This set of parameters is used on an independently simulated validation dataset with 5000 observations to find a reliable estimate of the true prediction performance. In the case studies we apply repeated double cross-validation (rdCV). This strategy, that is suitable for small datasets, selects the optimal parameter based on multiple repetitions instead of a single double cross-validation that can be optimistic or pessimistic \cite{filzmoser2009repeated}. Here, double (k-fold) cross-validation (dCV) is preferred above single k-fold CV, Monte Carlo CV (MCCV) or leave-one-out CV (LOOCV). Primarily because dCV is able to simultaneously provide an estimate for the prediction error and the tunable parameter, whereas single k-fold cross-validation only succeeds to perform one of these goals \cite{smit2007assessing}. Secondly, dCV has a reduced computational complexity compared to LOOCV. 

The remainder of this article is structured as follows. In the next section, both the proposed and selected traditional classification methods are formulated mathematically. Moreover, a description on the applied performance measures and cross-validation techniques is presented. In the section 'Simulation study' a detailed description of the design of the simulation study is provided, followed by the corresponding results. In the section 'Case studies', the major depression disorder (MDD) dataset and trisomy dataset are presented. Here, we first describe the design of the study and then present the results of the different prediction methods. The last section contains the discussion.

\section{Methods}
In this section we assume that $y_i$ denotes the group (or disease) indicator for subject $i =1,\ldots,n$ with $y_i=0$ a healthy control and $y_i=1$ a case. The (continuous) value of the $k^{\text{th}}$ biomarker for subject $i$ is denoted by $x_{i,k}$, where $k=1,\ldots,r$ and $r$ the number of observed biomarkers.

\subsection{Quantile based prediction}
Quantile based prediction (QBP) is a binary prediction method for continuous biomarkers, that uses the left and right tails of the empirical biomarker distributions of two groups to discriminate between cases and controls. QBP is able to discriminate when the tails of two groups are shifted with respect to each other (irrespective of mean differences or the remainder part of the distribution). The stronger the shift in the tails of a biomarker, the more likely it is that this shift is due to the disease. By combining multiple biomarkers a subject's total disease score can be constructed. This disease score represents some likelihood of being a case or control. 

The remainder of this paragraph follows the structure of QBP - that distinguishes the definition of its characteristics, the scoring mechanism based on these characteristics and the attribution of scores to individual subjects. An artificial example of a single biomarker $k$ is presented to illustrate the construction of the QBP characteristics (Figure~\ref{fig:DescriptionQBP} and Table~\ref{tab:exampleQBP}) and the scoring mechanism (Table~\ref{tab:exampleQBP.scoringmech}). Lastly, the arbitrary situation in Table~\ref{tab:exampleQBP.scoringmech.indiv} exemplifies the attribution of scores to a set of individuals in case of multiple biomarkers.

\subsubsection{QBP characteristics}
\begin{figure}[H] 
 \centering
  \includegraphics[width=.95\linewidth]{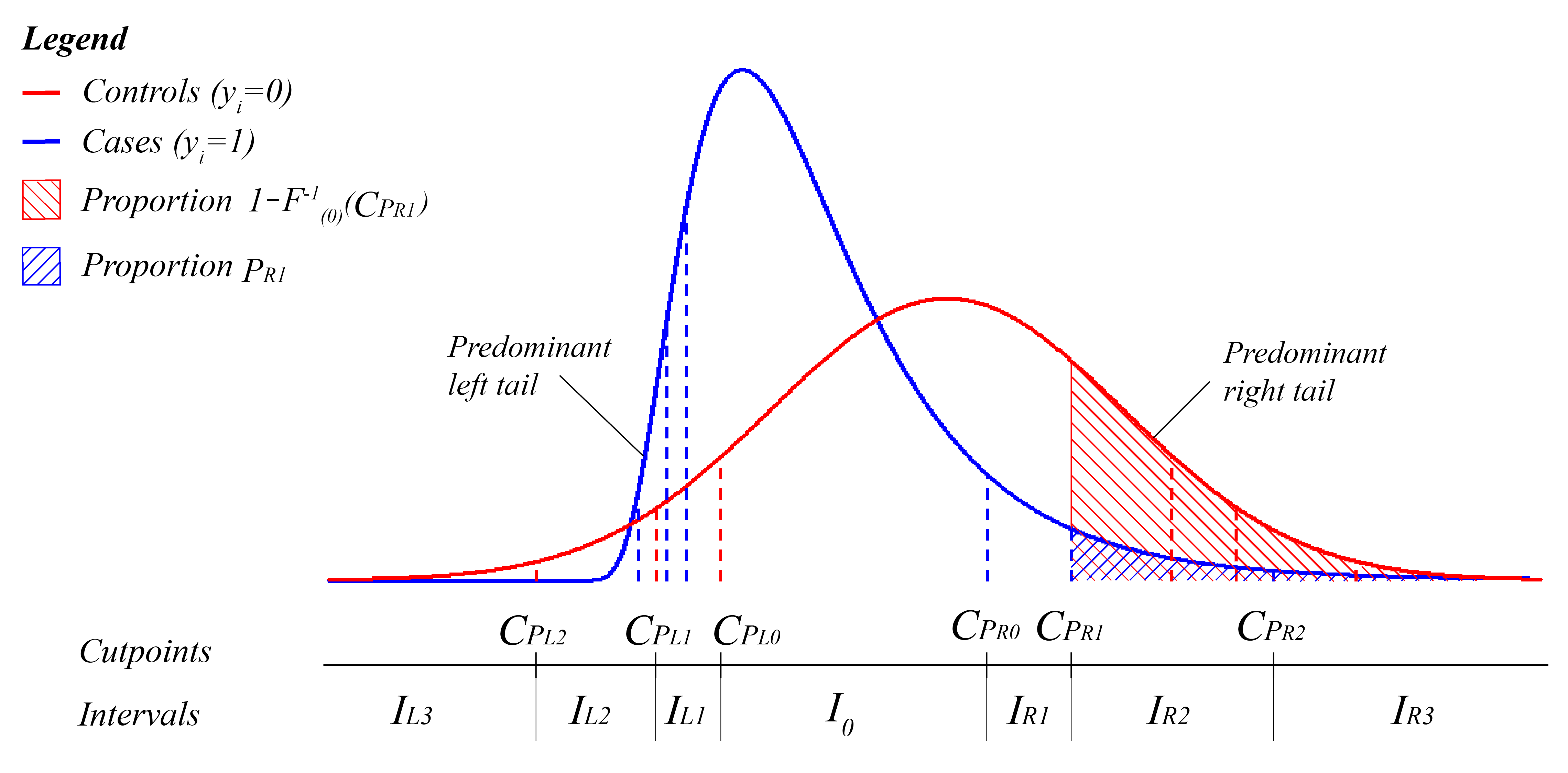}
  \caption{Illustration of QBP characteristics on data of a single biomarker $k$ (index $k$ suppressed)}
  \label{fig:DescriptionQBP}
\end{figure}

\begin{table}[H] 
\small
\centering
\caption{QBP characteristics on an arbitrary example using three ($m=2$) proportions per tail ($p_{L} = (p_{L_0},p_{L_1},p_{L_2})=(0.1,0.05,0.01)$ and ($p_{R} = (p_{R_0},p_{R_1},p_{R_2})=(0.9,0.95,0.99)$) for a single biomarker $k$ (index $k$ suppressed). Note that}
\label{tab:exampleQBP}
\begin{tabular}{lccccccc}
                    & $q_{p_{L_2}}$   & $q_{p_{L_1}}$    & $q_{p_{L_0}}$   &    & $q_{p_{R_0}}$   & $q_{p_{R_1}}$   & $q_{p_{R2}}$   \\
\hline
Percentiles ($y_i=0$)  & 273 & 372  & 424 &    & 796 & 849 & 947 \\
Percentiles ($y_i=1$)  & 357 & 380  & 396 &    & 644 & 713 & 880 \\
Predominant group   &  &   & $D_L=1$&    & $D_R=0$  &  &  \\
                    &       &        &       &    &       &       &       \\
                    & $C_{p_{L_2}}$  & $C_{p_{L_1}}$   & $C_{p_{L_0}}$  &    & $C_{p_{R_0}}$  & $C_{p_{R_1}}$  & $C_{p_{R_2}}$  \\ 
\hline
Cutpoints           & 273 & 372  & 424 &    & 644 & 713 & 880 \\
\\
                    & \multicolumn{3}{c}{$\mathbf{F_{(y_i)}^{-1}(C_{p_{L_s}})}$}  &    & \multicolumn{3}{c}{$\mathbf{1-F_{(y_i)}^{-1}(C_{p_{R_s}})}$}  \\ 
                   &  $p_{L_2} $ & $p_{L_1} $ & $p_{L_0}$ &  &$p_{R_0}$ &$ p_{R_1}$ &  $p_{R_2}$ \\                  \cline{2-4}  \cline{6-8}
Tail area ($y_i=0$)    & 0.01 & 0.05 & 0.1 &    & 0.407 & 0.240 & 0.03 \\
Tail area ($y_i=1$)    & 0.00 & 0.031 & 0.225 &     & 0.1   & 0.05  & 0.01  \\
\\
                    & $R_{p_{L_2}}$  & $R_{p_{L_1}}$   & $R_{p_{L_0}}$  &    & $R_{p_{R_0}}$  & $R_{p_{R_1}}$  & $R_{p_{R_2}}$  \\
\hline
Exceedratio         & 0   & 0.62  & 2.25  &    & 4.07  & 4.8  & 3   \\
\end{tabular}
\begin{tabular}{lccccccc}
\\
           & $I_{L_3}$   & $I_{L_2}$    & $I_{L_1}$   & $I_{0}$ & $I_{R_1}$   & $I_{R_2}$   & $I_{R_3}$ \\
\hline
Intervals           & $(-\infty,273]$   & $(273,372]$    & $(372,424]$   & $(424,644)$ & $[644,713)$   & $[713,880)$   & $[880,\infty)$  
\end{tabular}
\end{table}

The first step is to select a quantile (or percentile) $q_p$, with corresponding proportion $p$. For the left-tail percentile we select proportion $p_{L_0} < 0.50$ and we select the right-tail percentile with proportion $p_{R_0}>0.5$. Without loss of generality, we select the tail proportion $p_{R_0}$ based on symmetry such that $p_{R_0}=1-p_{L_0}$. The corresponding percentiles for the controls and cases for each biomarker $k$ are used to determine the predominant group in the left tail $D_{L,k} \in \{0,1\}$ and in the right tail $D_{R,k} \in \{0,1\}$. For each biomarker this is defined by
\begin{align}
D_{L,k}= \begin{cases} 
	0 & \text{ if } q_{p_{L_0},k}^{(0)} < q_{p_{L_0},k}^{(1)} \\ 
	1 & \text{ if } q_{p_{L_0},k}^{(0)} > q_{p_{L_0},k}^{(1)} \\ 
	\text{NA} & \text{ if } q_{p_{L_0},k}^{(0)} = q_{p_{L_0},k}^{(1)} 
 \end{cases}
,\qquad
D_{R,k}= \begin{cases} 
	0 & \text{ if } q_{p_{R_0},k}^{(0)} > q_{p_{R_0},k}^{(1)} \\ 
	1 & \text{ if } q_{p_{R_0},k}^{(0)} < q_{p_{R_0},k}^{(1)} \\ 
	\text{NA} & \text{ if } q_{p_{R_0},k}^{(0)} = q_{p_{R_0},k}^{(1)} 
 \end{cases} ,
\end{align}
with $q_{p,k}^{(0)}$ and $q_{p,k}^{(1)}$ the $p^\text{th}$ percentile ($p \in \{p_{L_0},p_{R_0}\}$) of group 0 (healthy control) and group 1 (cases) of biomarker $k$, respectively. Thus the predominant group has its percentile at proportion $p_{L_0}$ or $p_{R_0}$ more extreme than the other group. For example, in the illustration of QBP in Figure~\ref{fig:DescriptionQBP}, the control group ($y_i=0$) is predominant in the right tail and the case group ($y_i=1$) is predominant in the left tail.\\

In the second step the tails of the biomarkers that have a predominant group will be included in the discrimination of groups using scores. The tails having no predominant group ($D_{L,k}=\text{NA}$ or $D_{R,k}=\text{NA}$) are eliminated in the discrimination of groups by attributing a neutral score (value 0).

The third step is to define $m$ additional percentiles that are located further in the tail. The left and right tail now contain $m+1$ percentiles, with proportions $p_{L} = (p_{L_0},p_{L_1},\ldots,p_{L_m})$ in the left tail ($p_{L_{s-1}}>p_{L_{s}}$) and $p_{R} = (p_{R_0},p_{R_1},\ldots,p_{R_m})$ in the right tail. Again, without loss of generality, we use symmetry of the tails and take $p_{R_s}=1-p_{L_s}$. The cutpoints $C_{p,k}$ on biomarker $k$ for proportions $p\in\{p_L,p_R\}$ will be defined by the quantiles of the non-predominant group. In particular, for $s =1,\ldots,m$
\begin{align}
C_{p_{L_s},k}= q_{p_{L_s},k}^{(1-D_{L,k})} ,\qquad C_{p_{R_s},k}= q_{p_{R_s},k}^{(1-D_{R,k})}. 
\end{align}

With these cutpoints, we define $m+1$ intervals $I_{s,k}$ in each tail that will later be used to attribute scores to subjects. We define the intervals $I_{s,k}$ as follows
\begin{align}
 I_{L_s,k} = ( C_{p_{L_{s+1}},k},C_{p_{L_s},k}], \qquad I_{0,k} = ( C_{p_{L_0},k},C_{p_{R_0},k} ), \qquad   I_{R_s,k} = [ C_{p_{R_s},k},C_{p_{R_{s+1}},k} )  
 \label{eq:intervals}
\end{align}
with $s =1,\ldots,m$, $C_{p_{L_m+1},k}=-\infty$,  $C_{p_{R_m+1},k}=\infty$. In Figure~\ref{fig:DescriptionQBP}, the cutpoints and intervals of QBP are shown for an arbitrary biomarker.

The fourth step is to determine the exceedratio $R_{p_s,k}$ based on the cutpoints. Here, an exceedratio is a measure for the relative difference of mass in the tails of the predominant and non-predominant group. The higher the exceedratio at a cutpoint, the higher the probability that a new subject contained in this tail belongs to the predominant group. Note that the predominant group may be different for the left and the right tail and the predominant group has more mass in the tail at the $C_{p_{L_0},k}$ and $C_{p_{R_0},k}$ than the non-predominant group. Thus the exceedratio $R_{p_0,k}$ is greater than 1 at the corresponding quantile $q_{p_{L_0},k}^{(1-D_{L,k})}$ and $q_{p_{R_0},k}^{(1-D_{L,k})}$. However, this may not necessarily be greater than 1 for the other percentiles further in the tails. For the left and the right tail, the exceedratio is defined by
\begin{align}
R_{p_{L_s},k} = F_{(D_{L},k)}^{-1}(C_{p_{L_s},k})/p_{{L_s},k},	\qquad
R_{p_{R_s},k} = (1-F_{(D_{R},k)}^{-1}(C_{p_{R_s},k}) /(1-p_{{R_s},k}),
\label{eq:exceedratios}
\end{align}
with $F_{(0,k)}$ and $F_{(1,k)}$ the empirical distribution function of biomarker $k$ for the controls and the cases, respectively, and, $F^{-1}$ is the inverse function of $F$.

\subsubsection{Scoring mechanism}
Aiming to discriminate cases from controls, we will attribute the interval scores $V_{s,k}\in$ $\{ V_{0,k}$, $ V_{L_s,k}$, $ V_{R_s,k}\}$ to the different intervals $I_{s,k}\in$ $\{ I_{0,k}$, $ I_{L_s,k}$, $ I_{R_s,k}\}$, that were defined in \eqref{eq:intervals}, respectively. The result of the scoring mechanism - as explained below - applied on the artificial example from Figure~\ref{fig:DescriptionQBP} is shown in Table~\ref{tab:exampleQBP.scoringmech}.

Firstly, the predominant group in a tail will determine the sign of the interval scores. Whereas negative signs correspond to predominance of the healthy control group ($D_{L,k}=0$ or $D_{R,k}=0$), positive signs belong to predominance of the cases ($D_{L,k}=1$ or $D_{R,k}=1$). 

Secondly, to guarantee the predominant group has more mass in the tail for a certain percentile than the non-predominant group, and therefore a certain discriminating power, we introduce lower boundaries $R^*=(R_1^*, \ldots,R_m^*)$ on the exceedratios in \eqref{eq:exceedratios} with $R_s^*>1, \; \forall s \in\{1,\ldots,m\}$. To indicate whether these lower boundaries -- which we can choose ourselves -- are met for biomarker $k$, we apply binary exceedscores for the left-tail $e_{L,k}=(e_{L_0,k},\ldots,e_{L_m,k})$ and right-tail $e_{R,k}=(e_{R_0,k},\ldots,e_{R_m,k})$. Note that this can vary per tail (percentile) and biomarker, as can be seen in the artificial example in Table~\ref{tab:exampleQBP.scoringmech}. The binary exceedscores are mathematically defined by
\begin{align}
 e_{L_s,k} = \mathbbm{1}(R_{p_{L_s},k} \geq R_s^*), \qquad
 e_{R_s,k} = \mathbbm{1}(R_{p_{R_s},k} \geq R_s^*),
\end{align}
for $s=0,\ldots,m$ and with $\mathbbm{1}(A)$ an indicator value being $1$ if $A$ is true and zero otherwise. Note that for $s=1,\ldots,m$, the binary exceedratios $e_{L_{s-1},k}$ and $e_{R_{s-1},k}$ correspond to the intervals $I_{L_s,k}$ and $I_{R_s,k}$, respectively. 

Thirdly, intending to put more emphasis on subjects having (extreme) values in tails, we introduce maximal interval scores $v=(v_1,\ldots,v_m )$ such that $v_1\leq v_2\leq \ldots \leq v_m$. By appending these scores with the binary exceedratios, we will ensure that scores are only assigned in case of a certain discriminating power of a tail. For $s=1,\ldots,m$ we obtain the interval scores
\begin{equation}
\begin{split}
V_{L_s,k} & = (-1)^{(1-D_{L_k})} \cdot \max \{ v_1 \cdot e_{L_0,k},\ldots, v_s \cdot e_{L_{s-1},k} \},  \\
V_{R_s,k} & = (-1)^{(1-D_{L_k})} \cdot \max \{ v_1 \cdot e_{R_10,k},\ldots, v_s \cdot e_{R_{s-1},k} \}.
\end{split} 
\end{equation} 
Note that for increasing $s$, the functions $\max \{ v_1 \cdot e_{L_0,k},\ldots, \allowbreak v_s \cdot e_{L_{s-1},k}\}$ and $\max \{ v_1 \cdot e_{R_0,k},\ldots, v_s \cdot e_{R_{s-1},k}\}$ are non-decreasing and that the central interval $I_{0,k}$ always obtains a neutral score $V_{0,k} = 0$.

\begin{table}[H]
\small
\centering
\caption{Scoring mechanism for the arbitrary example in Table~\ref{tab:exampleQBP} using lower boundaries for the exceedratios $R^*=(2,3,5)$ and maximal interval scores $v=(v_1,v_2,v_3)=(1,2,3)$ for a single biomarker $k$ (index $k$ suppressed). Note that $D_L=1$ and $D_R=0$.}
\label{tab:exampleQBP.scoringmech}
\begin{tabular}{lccccccc}
                    & $R_{p_{L_2}}$  & $R_{p_{L_1}}$   & $R_{p_{L_0}}$  &    & $R_{p_{R_0}}$  & $R_{p_{R_1}}$  & $R_{p_{R_2}}$  \\
 \hline
Exceedratio         & 0   & 0.62  & 2.25  &    & 4.07  & 4.8  & 3   \\

Lower boundaries on exceedratio ($R^*$) & 5   & 3  & 2  &    & 2  & 3  & 5   \\
\\
\textbf{Intervals}           & $\mathbf{I_{L_3}}$   & $\mathbf{I_{L_2}}$    & $\mathbf{I_{L_1}}$   & $\mathbf{I_{0}}$ & $\mathbf{I_{R_1}}$   & $\mathbf{I_{R_2}}$   & $\mathbf{I_{R_3}}$   \\
\hline
           & $e_{L_2}$   & $e_{L_1}$    & $e_{L_0}$   &  & $e_{R_0}$   & $e_{R_1}$   & $e_{R_2}$   \\
Binary exceedscores & 0     & 0      & 1     &    & 1     & 1     & 0     \\
\hdashline
           & $v_3$   & $v_2$    & $v_1$   &  & $v_1$   & $v_2$   & $v_3$   \\
Maximal interval scores ($v$)    & 3     & 2      & 1     &   & 1     & 2     & 3     \\
\hdashline
           & $V_{L_3}$   & $V_{L_2}$    & $V_{L_1}$   & $V_{0}$ & $V_{R_1}$   & $V_{R_2}$   & $V_{R_3}$   \\
Interval scores & 1     & 1      & 1     & 0  & -1     & -2     & -2    \\
\end{tabular}
\end{table}

\subsubsection{Scoring individual subjects}
Now all elements of the QBP are determined, the disease scores ($DS_{i,k}$) can be computed for each subject $i$ per biomarker $k$. The disease score $DS_{i,k}$ is in essence a measure of the position of the biomarker value $x_{i,k}$ with respect to the predominant group. In order to prioritize specific biomarkers above others, biomarker weights $w=(w_1,\ldots,w_r)$ are introduced. The disease score $DS_{i,k}$ defined by
\begin{align}
DS_{i,k}= \begin{cases} 
	V_{L_s,k} \cdot w_k & \text{ if } x_{i,k} \in I_{L,s,k}, \\ 
	0  & \text{ if } x_{i,k} \in I_{0,k}, \\
	V_{R_s,k} \cdot w_k & \text{ if } x_{i,k} \in I_{R,s,k}, 
 \end{cases} 
\end{align}
with $s=1,\ldots,m$. Note that $x_{i,k}$ will always fall in one of the intervals $I_{L_m,k}$ $I_{L_{m-1},k}\ldots,I_{L_1,k}$, $I_{0,k}$, $I_{R_1,k},\ldots,I_{R_{m-1},k},I_{R_m,k}$. By summing over all biomarkers, a total disease score $TDS_i=\sum_{k=1}^m DS_{i,k}$ per subject $i$ can be calculated. An extreme positive value for subject i, indicates that the subject is most likely a case, while an extreme negative value means that subject $i$ most likely a control. A value of zero would indicate that the subject is as likely a case as a control. This procedure is applied on an arbitrary example in Table~\ref{tab:exampleQBP.scoringmech.indiv}.

\begin{table}[h!]
\centering
\small
\caption{Arbitrary example of the calculation of the summation of disease scores $DS_{i,k}$ and the total disease score $TDS_i$ for subjects $i\in \{a,b,c\}$. Here, we have the biomarker weights $w=(1,1,1,1,1)$ and maximal interval scores $v=(v_1,v_2,v_3)=(1,2,3)$}
\label{tab:exampleQBP.scoringmech.indiv}
\begin{tabular}{lcccccccc}
\toprule
         & \multicolumn{5}{c}{Biomarker $k$}     &  					& Subject  & TDS  \\ \cline{2-6}
Interval & $1$         & $2$  & $3$ & $4$ & $5$  &           	&      &  \\
\midrule
$I_{L_3,k}$      & $1$         & $-1$ 		& $2$ 		& $-3^{c}$ 	& $-3$ 		&          &   a  & $3$ \\
$I_{L_2,k}$      & $1^{a}$     & $-1$ 		& $2$ 		& $-1$ 		& $-2^{c}$ 	&          &   b  & $0$ \\
$I_{L_1,k}$      & $1$         & $-1^{c}$ 	& $1$ 		& $-1^{b}$ 	& $0$  		&          &   c  & $-7$ \\
$I_{0,k}$      	 & $0^{b}$     & $0^{a}$  	& $0^{a,c}$	& $0$  		& $0^{b}$  	&          &      &  \\
$I_{R_1,k}$      & $-1^{c}$    & $0^{b}$  	& $1^{b}$ 	& $0$  		& $0$  		&          &      &  \\
$I_{R_2,k}$      & $-2$        & $0$  		& $2$ 		& $0^{a}$  	& $2^{a}$  	&          &      &  \\
$I_{R_3,k}$      & $-2$        & $0$  		& $3$ 		& $3$  	& $2$  		&          &      & \\
\bottomrule
\end{tabular}
\end{table}

\subsection{(Penalized) Logistic regression} 
As described in \textcite{hosmer2000introduction}, logistic regression considers $n$ independent observations $\{(y_i,x_i);i=1,\ldots,n\}$, where $y_i$ corresponds to a disease ($y_i=1$) or no disease ($y_i=0$) and $\bm{x}_i=(1,x_{i,1},\ldots,x_{i,r})$ is the vector of independent predictor variables, which are the results of the $r$ biomarkers. The logistic regression model assumes that,
\begin{equation}
\mathbb{P}(Y_i=1|\bm{x}_i)=\pi(\bm{x}_i)=1-\mathbb{P}(Y_i=0|\bm{x}_i)),
\label{eq:logitmodel}
\end{equation}
with $Y_i$ Bernoulli$(\pi(\bm{x}_i))$ distributed and $\pi(\bm{x}_i)$ given by
\begin{equation}
\pi(\bm{x}_i) = \frac{\exp(\beta_0+ \sum_{j=1}^r x_{i,j} \beta_j)}{1+\exp(\beta_0 + \sum_{j=1}^r x_{i,j} \beta_j))} 
\label{eq:logitfunction}
\end{equation}
In case the number of events is large enough to be able to estimate all model parameters, maximum likelihood estimators can be used. The log-likelihood function for $\bm{y}=(y_1,\ldots,y_n)$ is given by
\begin{equation}
l(\beta)=  \sum_{i=1}^n [y_i \cdot  \pi(\bm{x}_i)+(1-y_i) \cdot \log(1-\pi(\bm{x}_i))].
\label{eq:likelihoodfunction}
\end{equation}

In case the number of events is sparse, a penalized logistic regression can be used to determine the most promising or relevant biomarkers. The penalized logistic regression model maximizes the log-likelihood function $l(\beta)$ in \eqref{eq:likelihoodfunction} with a penalty term $P(\beta)$, i.e. maximizes $l_\lambda(\beta)= l(\beta) -\lambda P(\beta)$ over $\beta$ for a fixed value of $\lambda$ that determines the strength of the penalty. Three well known penalty functions are the lasso \cite{tibshirani1996regression}, elastic net (EN) \cite{zou2005regularization} and the ridge \cite{hoerl1970ridge} (see \eqref{eq:penalized.methods}).
\begin{equation}
\label{eq:penalized.methods}
\begin{aligned}
\text{Lasso: } P (\beta) = \sum_{k=1}^r |\beta_k| & \qquad \text{EN: } P_{\alpha} (\beta) = \sum_{k=1}^r [ \frac{1-\alpha}{2}  \beta_k^2 + \alpha |\beta_k|] & \qquad
\text{Ridge: } P (\beta) = \sum_{k=1}^r \beta_k^2  
\end{aligned}
\end{equation}
with $\alpha$ an additional parameter for the elastic net.

\subsection{Principal Components Logistic Regression}
First of all, we briefly describe the concept of principal component analysis (PCA) in line with a more comprehensive description of this method in \textcite{aguilera2006using}. Let all observations be contained in matrix $\bm{X}=(x_{i,k})_{n \times r}$, with column vectors $\bm{X}_1, \bm{X}_2, \ldots, \bm{X}_r$. Furthermore, denote the sample covariance matrix $\bm{S}=(s_{k,l})_{r \times r}$ with the elements $s_{k,l}=\tfrac{1}{n-1}\sum_{i=1}^n(x_{i,k}-\bar{x}_k)(x_{i,l}-\bar{x}_l)$, with sample means given by $\bar{x}_k=\tfrac{1}{n}\sum_{i=1}^n x_{i,k} $, with $k=1,\ldots,r$. In order to simplify, without loss of generality, it is considered that the observations are centered, so that $\bar{x}_1=\ldots=\bar{x}_r=0$, and the sample covariance matrix $\bm{S}=(s_{k,l})_{r \times r}=\tfrac{1}{n-1} \bm{X'}\bm{X}$. 

The sample principal components (pc's) are defined as orthogonal linear spans with maximum variance of the column matrix $\bm{X}$, denoted by $Z_k=\bm{XV}_k$ with $k=1,\ldots,r$. The vectors $\bm{V}_1,\ldots,\bm{V}_r$ that define the pc's, are the eigenvectors of the sample covariance matrix $\bm{S}$ associated to their corresponding eigenvalues $\lambda_1\geq \ldots \geq \lambda_r\geq 0 $. These eigenvalues are again the variances of the corresponding pc's. If we denote by $\bm{Z}$ the matrix whose columns are the sample pc's, it can be expressed as $\bm{Z}=\bm{XV}$, with $\bm{V}=(v_{k,l})_{r \times r}$ being the matrix whose columns are the eigenvectors of the sample covariance matrix.
Note that the sample variance can be decomposed as $\bm{S}=\bm{V \Delta V}'$, with $\bm{V}$ orthogonal, $\bm{V}'$ being the transposed of $\bm{V}$ and $\bm{\Delta}=diag(\lambda_1,\ldots,\lambda_r)$, so the matrix of observations is given by $\bm{X}=\bm{ZV'}$. This pc decomposition has given us an approximate reconstruction of each original observation in terms of a reduced number of pc's that was selected based on explained variance, namely
\begin{equation}
\bm{X}_k=\sum_{l=1}^s \bm{Z}_l v_{k,l}, \; k=1,\ldots,r, \; \text{with } s\leq r.
\label{eq:reducedpc}
\end{equation}
The percentage of the variability that is accounted for by the model is given by
\begin{equation}
\frac{\sum_{l=1}^s \lambda_l \cdot 100}{\sum_{l=1}^r \lambda_l}, \; \text{with } s\leq r.
\label{eq:reducedpcvariabilitypercentage}
\end{equation}
Now that the pc's are obtained, the logit model is applied, with \eqref{eq:logitfunction} being replaced by
\begin{equation}
\pi_s(\bm{Z}_i) = \frac{\exp\{\beta_0 + \sum_{k=1}^s\sum_{l=1}^s z_{i,l}v_{k,l}\beta_k\}}{1+\exp\{\beta_0 + \sum_{k=1}^s\sum_{l=1}^s z_{i,l}v_{k,l}\beta_k \}}=
\frac{\exp\{\beta_0 + \sum_{l=1}^s z_{i,l}\gamma_l \}}{1+\exp\{\beta_0 + \sum_{l=1}^s z_{i,l}\gamma_l \}}
\label{eq:logitmodelpcs}
\end{equation}
with $z_{i,l}$ being the elements of the pc's matrix $Z=XV$ and $\gamma_l=\sum_{k=1}^r v_{k,l}\beta_k,\; k=1,\ldots,r$.

\subsection{Linear Discriminant Analysis}
\label{sec:LDA}
In the search for a separating hyperplane using linear discriminant analysis (LDA), two approaches can be distinguished, namely LDA based on the Bayes' rule and Fisher-LDA. We focus on Bayesian LDA, since it appears to be more suitable with a large number of covariates \cite{vera2011discrimination}.

As extensively described in \textcite{friedman2001elements}, Bayesian LDA assumes Gaussian class densities with a common covariance matrix for all classes. For the binary case, this comes down to observing the log-ratio of the cases ($y=1$) and the controls ($y=0$). This log-ratio is defined by

\begin{equation}
\log \frac{P(Y=1|X=x)}{P(Y=0|X=x)}=\log \frac{\pi_1}{\pi_0}-\frac{1}{2}(\mu_1+\mu_0)^T\Sigma^{-1}(\mu_1-\mu_0)+x^T\Sigma^{-1}(\mu_1-\mu_0),
\end{equation}
with the prior distributions $\pi_1$ and $\pi_0$ and the mean vectors of the multivariate Gaussian $\mu_1$ and $\mu_0$ of the cases and controls, respectively. In addition, $\Sigma^{-1}$ denotes the common covariance matrix and $x$ the vector of biomarker values of a subject.

\subsection{Partial Least Squares - Linear Discriminant Analysis}
Partial least squares (PLS) \cite{wold1985partial} was first introduced for a continuous response, however, later a two-step approach for binary classification was proposed, namely PLS-LDA \cite{nguyen2002tumor}. Here, PLS is used for dimension reduction and then (Fisher)-LDA is used on the PLS latent variables. The underlying idea of PLS regression is to find uncorrelated linear transformations of the original predictor variables which have high covariance with the response variables. In this case, the classes of cases and controls are represented as binary responses and treated as if they were continuous in the projection on the latent structure of PLS \cite{boulesteix2004pls}. Since the principle of LDA is already explained in Subsection~\ref{sec:LDA}, we will now explain the PLS dimension reduction using the SIMPLS algorithm \cite{de1993simpls}.

Let us first recall that $\bm{X}\in \mathbb{R}^{n\times r}$ denotes the matrix containing all biomarker observations. Then, $\bm{Z}=\bm{XA}\in \mathbb{R}^{n\times s}$ denotes the matrix of linear transformations, with the column vectors $\bm{Z}_1,\ldots,\bm{Z}_s$ representing the PLS latent variables of $\bm{Z}$. Here, the matrix $\bm{A}\in \mathbb{R}^{r\times s}$ defines the linear transformation and contains the vectors $a_1,\ldots,a_s$ as its columns. The SIMPLS algorithm determines the vector $a_1,\ldots,a_s$ by computing linear transformations of $X$ and linear transformations of $y=(y_1,\ldots,y_n)$ which have maximal covariance, under the constraint that the linear transformations of $X$ (the PLS latent variables) are mutually uncorrelated. In particular, we first determine the unit vector $a_1$ and scalar $b_1$ maximizing the empirical covariance $\hat{COV}(X a_1,b_1 y )$. Then for all $l=2,\ldots,s$, the unit vector $a_l$ and scalar $b_l$ maximize $\hat{COV}(X a_l,b_l y )$ subject to $\hat{COV}(X a_l,b_u y)=0$ for all $u=1,\ldots,l-1$. Note that before applying the SIMPLS algorithm, $y$ and the columns of $X$ need to be centered.

Now we have obtained the matrix $\bm{Z}=\bm{XA}$, Fisher LDA is applied using $Z_1,\ldots,Z_s$ as predictor variables. In order to determine the optimal number of components $s$ that results into the best classification performance, cross-validation (see \ref{sec:crossvalidation}) is performed.

\subsection{Support Vector Machine}
Support vector machine (SVM) is a generalization of optimal separating hyperplanes to the non-separable case and creates non-linear decision boundaries for classification composed by taking linear combinations of a largely transformed (sometimes infinite) version of the feature space \cite{boser1992training,cortes1995support}. 
 
In both the separable and non-separable situation, we have $n$ independent observations $\{(y_i,x_i);i=1,\ldots,n\}$, where $y_i$ corresponds to a disease ($y_i=1$) or no disease ($y_i=-1$) and $\bm{x}_i=(1,x_{i,1},\ldots,x_{i,r})$. Here, we can define a hyperplane by
\begin{equation}
\{\bm{x}: f(\bm{x}) =\bm{x}^T\beta+\beta_0=0\}
\end{equation}
and the classification rule $\text{sign}[\bm{x}^T\beta+\beta_0]$ to distinguish between cases and controls.

As extensively described in \textcite{friedman2001elements}, we can find the optimal separating hyperplane that maximizes the margin (M) between the cases and controls, by solving the following optimization problem
\begin{equation}
\label{eq:svmseparable}
\begin{aligned}
\max_{\beta,\beta_0,||\beta||=1} & \quad M \\
\text{subject to} & \quad y_i(\bm{x_i}^T\beta+\beta_0)>M, \;  i=1,\ldots,n.
\end{aligned}
\end{equation}
Note that the problem \eqref{eq:svmseparable} can not be solved for the non-separable case. To allow for overlap in the feature space between cases and controls, a slack variable $\xi_i$ that allows for points on the wrong side of the decision boundary was introduced. This concept of accepting errors in the training set is called soft margin. When extending \eqref{eq:svmseparable} with this slack variable that is proportional to the margin, we the following optimization problem
\begin{equation}
\label{eq:svmoptimfn}
\begin{aligned}
\min_{\beta,\beta_0} & \quad \frac{1}{2}||\beta||^2+C \sum_{i=1}^{n}\xi_i \\
\text{subject to} & \quad \xi_i\geq0, \; y_i(\bm{x_i}^T\beta+\beta_0)>1-\xi_i, \;  \forall_i,
\end{aligned}
\end{equation}
where the parameter $C$ is a cost parameter that can be used for regularization \cite{cortes1995support}. Note that for $C=\infty$, we obtain the separable case again.

The quadratic optimization problem can be rewritten as a dual SVM problem, such that it only depends inner products. We obtain
\begin{equation}
\label{eq:svmoptimdual}
\begin{aligned}
\max_{\alpha} & \quad  \sum_{i=1}^{n}\alpha_i - \frac{1}{2}\sum_{i=1}^{n}\sum_{j=1}^{n}\alpha_i\alpha_{j}y_iy_{j}\bm{x_i}^T\bm{x_{j}} \\
\text{subject to} & \quad 0\leq \alpha_i\leq C
\end{aligned}
\end{equation}
This form makes it possible to apply the kernel trick, in which the inner product $\bm{x_i}^T\bm{x_j}$ is replaced by $K(\bm{x_i},\bm{x_j})$ representing a kernel that enlarges the original feature space using polynomials or splines. The main advantage of this enlarged space is the enhanced training-class separation. To avoid over-fitting, one can make a trade-off between model complexity and error frequency by changing the soft margin cost parameter $C$ \cite{cortes1995support}.

In this study, we apply two types of kernels, namely the linear $K(\bm{x_i},\bm{x_j})=\langle h(\bm{x_i}),h(\bm{x_j})\rangle$ and the radial base function (RBF) with $K(\bm{x_i},\bm{x_j})=\exp(-\gamma ||\bm{x_i}-\bm{x_j}||^2), \gamma>0$.  

\subsection{Random Forest}
A random forest algorithm is an ensemble of individual regression (or decision) trees, that can be used for both regression or classification problems. 

Each individual tree is grown by recursively selecting a number of random features from the training sets composed of bootstrap samples from the original data, and consequently creating two daughter nodes at the feature that provides the best split. Here, the best split is defined such that the response can be predicted in the best possible way. This partitioning at nodes continues until a stopping criterion has been met. In the end, each tree provides a tree-structured classifier $\hat{C}_b(x)$. Since individual trees have a relatively low bias but are noisy, it is beneficial to average individual trees to reduce the variance \cite{friedman2001elements}.

The random forest classifier consists of a majority vote of the collection of all individual tree classifiers. In specific, 

\begin{equation}
\begin{aligned}
\hat{C}_{\text{rf}}^B(x)=\text{majority vote} \{ \hat{C}_b(x)\}_1^B.
\end{aligned}
\end{equation}

By combining all votes of $x$ for which $x$ is not contained in the training set we obtain the out-of-bag classifier of input $x$ \cite{friedman2001elements}. The proportion of these out-of-bag votes is used to determine the classification performance in terms of AUC, as explained in Subsection~\ref{par:perf.measures}.

\subsection{k-Nearest Neighbors}
The philosophy behind k-nearest neighbors (kNN) is that observations that show a high degree of similarity are likely to share the same class label. Here, the distance between data points is considered a measure for similarity. The (kNN) technique searches, for each point in the validation dataset, the $k$ datapoints from the training set that are closest in terms of Euclidean distance. 

The classification is decided by majority vote, with ties broken at random. If there are ties for the kth nearest vector, all candidates are included in the vote \cite{friedman2001elements}. In the case of skewed class distributions, this majority voting might be somewhat problematic, since one class is dominant by default \cite{coomans1982alternative}.

Generally, larger values of $k$ make the classification less susceptible to the effect of noise \cite{everitt2011miscellaneous}. The value of value $k$ is based on cross validation, as explained in Subsection~\ref{par:perf.measures}. Moreover, in this study, we normalize all input variables before applying kNN.

\subsection{XGBoost}
EXtreme Gradient Boosting (XGBoost) is a variant of the Gradient Boosting Machines (GBM) algorithm that includes regularization and dedicates its name due to its highly efficient algorithmic implementation \cite{chen2016xgboost}. XGBoost is a machine learning technique that uses the boosting principle by combining weakly performing individual trees into an ensemble of trees representing a strong classifier. The primary purpose of boosting is to reduce bias, but also suitable for reducing variance \cite{zhou2012ensemble}.

XGBoost evaluates the classification performance in each iteration and aims to correct for the errors in each consequent step by adding a new tree. This new tree is trained on the gradient, that is determined by deriving the negative gradient of the loss function with respect to the predictions. The algorithm repeats this process for pre-specified number of iterations. Regularization is applied to avoid overly complex models. The predictions of the final ensemble of trees are the weighted sum of the predictions on the log odds scale from the individual tree models.

As extensively described in \textcite{chen2016xgboost}, XGBoost aims to minimize the regularized objective function
\begin{equation}
\mathcal{L}^{(t)}=\sum_{i=1}^n l\left(y_i,\hat{y_i}^{(t-1)}+f_t(x_i)\right)+\Omega(f_t),
\label{eq:xgboost}
\end{equation}

where $\hat{y_i}^{(t)}$ is the prediction of the $i$-th instance at the $t$-th iteration, $ \Omega (f)$ the regularization term. In each iteration, a new tree $f_t$ is added aiming to minimize \eqref{eq:xgboost}. Given the convex nature of the loss function $l$, a second order approximation of $\mathcal{L}^{(t)}$ is applied.

In addition to the regularization of weights leaf weights, shrinkage is implemented in the XGBoost algorithm by scaling newly added weights with a factor $\nu$, with $0 < \nu \leq 1$. Here, the lower the value for $\nu$, the higher the computation time. Empirically, it was found that small values ($\nu<0.1$) lead to much better generalization error \cite{friedman1999stochastic}.

\subsection{Performance measures}
\label{par:perf.measures}
To assess the performance of the classification of cases and controls for all methods, a receiver operating characteristic (ROC) curve is constructed by means of the sensitivity (true positive rate) and the specificity (1$-$false positive rate) using different cut-offs for the probability of an outcome \cite{steyerberg2010assessing}. Here, each method requires a different way to define these cut-offs. 

For QBP, we use the total disease score of each subject as different cut-offs. The logistic regression approaches naturally have an estimation of the class probabilities. Both LDA and PLS-LDA use the posterior probability that follows from the Bayesian way of modeling. SVM applies Platt-scaling to come up with the posterior probability for the classifier \cite{Platt99probabilisticoutputs}. The proportion of the votes is used for random forest and kNN. Lastly, XGB uses the \textit{'binary:logistic'} objective function to define the class probabilities.

For each cutpoint the sensitivity and specificity are defined by
\begin{equation}
\begin{split}
\text{\textit{Sensitivity}} = \frac{\text{\textit{TP}}}{\text{\textit{TP}} +\text{\textit{FN}}}, \qquad \text{\textit{Specificity}} = \frac{\text{\textit{TN}}}{\text{\textit{TN}} +\text{\textit{FP}}},
\end{split}
\label{eq:SensSpec}
\end{equation}
with $TP=\;$true positives, $TN=\;$true negatives, $FP=\;$false positives, $FN=\;$false negatives. In fact, the area under the ROC curve (AUC) represents the probability that a randomly chosen positive example is correctly rated (ranked) with greater likelihood than a randomly chosen negative example. Moreover, this probability of correct ranking is the same quantity estimated by the non-parametric Wilcoxon statistic \cite{bradley1997use}. Thus the higher the AUC the better the classification. Here, a perfect separation of cases and controls is denoted by $AUC=1$ and a separation which is not better than random is denoted by $AUC=0.5$. To determine the AUC, the trapezoidal integration method is used, which is implemented by the [R] software package \textit{'ROCR'} \cite{ROCRpackage}. 

The performance of the biomarker inclusion is evaluated with the sensitivity, specificity and accuracy. Here, the accuracy defined by \\
\begin{equation}
 \text{\textit{Accuracy}} = \frac{\text{\textit{TP}} + \text{\textit{TN}}}{\text{\textit{TP}} + \text{\textit{TN}} + \text{\textit{FP}} + \text{\textit{FN}}}.
\label{eq:SensSpec2}
\end{equation}
The closer the accuracy is to one the better the classification.\\

\subsection{Cross-validation}
\label{sec:crossvalidation}
A major difference between the simulation scenarios and the case studies is the (dis-)ability to generate datasets of an arbitrary size. Therefore, we choose to apply different cross-validation (CV) strategies for the simulation scenarios and the case studies.

For all simulation scenarios, we generate a total number of 500 repetitions, each with a separate training set of size $n$ and new validation set with 5000 subjects. Note that the training set size depends per simulation scenario, and is defined in Table~\ref{tab:SimulationSettings}. For every single repetition, we apply 6-fold CV on the training set to determine the optimal set of tunable parameters for a particular method. Here, the parameter settings with the highest mean AUC over all 6 folds are selected as the optimal set of tunable parameter $t_{opt}$. So $t_{opt} = \argmax_t \{ \text{mean}(\text{AUC}(t)) \}$. Then the predictive performance of each method is assessed on the validation set using the optimal parameters obtained via CV on the training data. 

In the case studies we apply repeated double CV with a total number of 500 repetitions. For each repetition, 6-fold outer-CV is applied to assess the predictive performance. Here, the dataset is divided into a training and validation (also called test) set, according to a 5:1 ratio. For all 6 permutations of the outer-CV training and outer-CV validation set, $t_{opt}$ is determined using 6-fold inner-CV. Consequently, this parameter is applied in the model fit on the full outer-CV training set and used to assess the predictive performance on the outer-CV validation set. Since one particular split of the outer-CV could skew the results positively or negatively, we use different splits per repetition to obtain an unbiased estimate of the predictive performance. This way of cross-validation is especially useful when limited data or just one dataset is available \cite{filzmoser2009repeated}. In addition, the prediction error is representative for new samples \cite{westerhuis2008assessment}.

\subsection{Tunable parameters}
In this study, the considered methods vary in the number of tunable parameters. Where LR and LDA have no tunable parameters, the methods PLR, PLS-LDA, PCLR and kNN just have a single tunable parameter. Lastly, QBP, RF, SVM and XGBoost use numerous tunable parameters.

In specific, we define the penalty term for PLR, the number of principal components for PLS-LDA and PCLR, and the number of neighbors for kNN. Here, the penalty term of PLR was obtained using the automated cross-validation procedure of the $glmnet$ package \cite{glmnet} of [R]. For both PLS-LDA and PCLR we selected the optimal number of sparse components via CV $ncomp\in\{1,\ldots,p\}$, with $p$ the number of covariates. For kNN, the optimal number of neighbors was selected from a set of candidates with step size from 1 to 20 and an increasing step size above 20 neighbors.

QBP has in principle many tunable parameters, but we made some decisions upfront. We fix both the number of percentiles and the corresponding proportion choice -- obtaining $\{q_1 , q_5 , q_{10} , q_{90} ,\allowbreak q_{95} , q_{99}\}$ -- and keep all biomarker weights equal. The settings that are determined by cross-validation are the lower boundaries of the exceedratios and the maximal interval scores, which are defined by the sets $R^*= (R_1^*,R_2^*,R_3^*) \in \{ (1.5,2,3)$, $ (1.5,2,5)$, $(1.5,2.5,5)$, $(1.4,2.5,8)$, $(2,3,6)$, $(2,3,10) \}$ and $v=(v_1,v_2,v_3) \in \{(1,2,3),(1,4,9)\}$ respectively. Eventually, the optimal setting is selected from $R^* \times v$.

To reduce the computational complexity for RF, XGBoost and SVM in the final simulation study, we have selected a subset of a larger grid of candidate tunable parameters. Each combination of tunable parameters was used to fit a model on a training dataset of 5000 subjects, after which the performance was evaluated on the corresponding validation datasets with 5000 subjects. To determine the final subset, we considered all scenarios and selected the most relevant tuning parameters using a regression approach.

For RF, checking the convergence of the out-of-bag error resulted in a total number of trees $ntree$ of 3000. In addition, we chose number of variables sampled randomly at each split to be $mtry\in\{6,9,12,15,18\}$. For XGBoost, the final set of tunable parameters is $nrounds=300$, $eta\in\{0.05,0.15,0.3\}$, $max\_depth\in\{2,4\}$, $colsample\_bytree=0.75$, $min\_child\_weight=2$, $gamma=0$ and $subsample=1$.

\section{Simulation study}

\subsection{Model and settings}
The group indicator $y_i \in \{0,1\}$ was divided such that we obtain $\phi \cdot n$ cases ($y_i=1$) and $(1-\phi) \cdot n$ controls ($y_i=0$), where $\phi$ denotes the proportion of cases and $n$ the total number of participants $n$. Then the variables $z_{i,1},\ldots,z_{i,r}$ were drawn from a multivariate distribution with mean $0$ and variance-covariance matrix R. In the statistical software [R], we used the \textit{mvrnorm} function of the \textit{'MASS'} package to create the variables $z$ \cite{MASSpackage}. Then the variables $v_{i,1},\ldots,v_{i,r}$ were taken equal to
\begin{align}
v_{i,k}=\mu_{i,k}+\sigma_{i,k}z_{i,k},
\end{align}
with $\mu_{i,k}=\alpha_k + \beta_k y_i$ and $\sigma_{i,k}=\eta_k \cdot (1+\nu_k - 2\nu_k y_i)$ for all $i=1,\ldots,n$ and $k=1,\ldots,r$. When $\beta_k=0$ and $\nu_k=0$, cases and controls are drawn from the same distribution and the variable $v_{i,k}$ does not contribute directly to the classification of cases and controls. Moreover, the variables $\alpha_k$ and $\eta_k$ differ per dataset and are based on (a transformation of) the MDD case study and correspond to its mean and standard deviation, respectively. Note that positivity of $\sigma_{i,k}$ is ensured in the simulation study by positivity of $\eta_k$ and selecting $\nu_k$ such that $-0.5 < \nu_k <0.5$ for all $k$. Finally, we take a transformation of the variables $v_{i,k}$ to have non-normally distributed variables that can be skewed. Thus, $x_{i,k}= \Psi_k(v_{i,k})$ with $\Psi_k$, the transformation that can be unique for each variable $k=1,\ldots,r$.

In total, 9 different types of datasets are simulated, with varying transformations, sample sizes and number of relevant biomarkers ($\beta_k \neq 0$ and/or $\nu_k\neq0$). We distinguish two types of transformations, namely
\begin{align}
\Psi_k(x) = x, \qquad \Psi_k(x) = \exp(x), 
\end{align}
where the former one results into normally distributed data and the latter in log-normally distributed data. We select only one type of transformation per dataset and biomarker, except for dataset 5, where for some covariates the biomarker distributions of the controls are normally distributed and those of the cases log-normally distributed, to create differences in terms of skewness. Here, the values for $\alpha_k$ and $\nu_k$ of $\Psi_{k_0}=I^*$ of the control group are chosen such its expected average and variance are similar to those of the distribution of the cases with $\Psi_{k_1}=\exp$.

The variance-covariance matrix R was always the same and based on the MDD case study in this paper. The full specification of R and the settings for $\alpha_k$ and $\eta_k$ are given in Table~\ref{tab:SimulationSettings}. The relevant biomarkers varied in number and in the way that they were different between cases and controls. Some varied only in mean ($\beta_k \neq 0$), some varied only in variance ($\nu_k \neq 0$) and other varied in both means and variances. A full overview of the choices is given in Table~\ref{tab:SimulationSettings}. Each dataset type is simulated 500 times.

Dataset 1,2 and 3 have the identity biomarker transformation and therefore obey a normal distribution. The datasets differ in terms of number of relevant biomarkers. Moreover, the applied linear transformation is a shift in mean $\beta_k$ of one standard deviation of that particular biomarker $\sigma_k$. Dataset 4 is also normally distributed with a shift in standard deviation $\nu_k$. A difference in skewness for some of the biomarkers is simulated in dataset 5. Whereas datasets 6 and 7 solely have log-normally distributed biomarkers, dataset 8 has a mixture of normally and log-normally distributed biomarkers. Dataset 6 to 8 show a fixed shift in mean $\beta_k$ and/or shift in standard deviation $\nu_k$. Besides that these datasets vary in the total number of participants $n$, where both a balanced and unbalanced number of cases and controls is considered. Except datasets 6c, 7c and 8c that consider an unbalanced setting with $\phi=1/5$, all other datasets are balanced ($\phi=1/2$).

\clearpage

\begin{table}[t!]
\small
  \centering
  \caption{Full design simulation study: All characteristics of all 9 datasets\\
  \small{\textit{Note that the transformation $\Psi_k = I$ equals $\Psi_k(x) = x$ and $\Psi_k = \exp$ equals $\Psi_k(x) = \exp(x)$. Moreover, $\alpha_k$ and $\eta_k$ denote the applied mean and standard variance derived from the MDD case study. Lastly, $\beta_k$ and $\nu_k$ denote the shift in mean and standard deviation. Lastly, empty cells correspond to a value of 0.}}}
    \label{tab:SimulationSettings}%
   \scalebox{0.70}{
    \begin{tabular}{l||ll|ll|ll||c:c:c:c:cc:c:c:ccc}
    \toprule
    \multicolumn{7}{c}{} & \multicolumn{4}{c}{\textbf{Datasets 1-4}} & \multicolumn{2}{c}{\textbf{Dataset 5}} & \multicolumn{2}{c}{\textbf{Dataset 6 to 7}}  & \multicolumn{3}{c}{\textbf{Dataset 8}} \\
    \multicolumn{7}{c}{} & \multicolumn{4}{c}{\textbf{($\Psi_k=I$)}} & \multicolumn{2}{c}{\textbf{($\Psi_{k_y}\in\{I^*,exp\}$)}} & \multicolumn{2}{c}{\textbf{($\Psi_k=exp$)}}  & \multicolumn{3}{c}{\textbf{($\Psi_{k}\in\{I,exp\}$)}} \\
\hline
 \multicolumn{1}{c||}{}  & \multicolumn{6}{c||}{} &&&&&&&&&&\\
 
\multicolumn{1}{c||}{}  & \multicolumn{6}{c||}{} & \textbf{1} & \textbf{2} & \textbf{3} & \textbf{4} &  \multicolumn{2}{c:}{\textbf{5}} &  \textbf{6} & \textbf{7} &   \multicolumn{3}{c}{\textbf{8}}\\

\multicolumn{1}{c||}{}  & \multicolumn{6}{c||}{} & $\beta_k=0$ & & &$\beta_k=0$ &  \multicolumn{2}{c:}{$\beta_k=0 $}&   & $\beta_k=0$ &   \multicolumn{3}{c}{}\\
\multicolumn{1}{c||}{}  & \multicolumn{6}{c||}{} & $\nu_k=0$ & $\nu_k=0$ & $\nu_k=0$ &  &  \multicolumn{2}{c:}{$\nu_k=0 $}&  $\nu_k=0$ &  &   \multicolumn{3}{c}{}\\
          
 \multicolumn{1}{c||}{}  & \multicolumn{6}{c||}{\textbf{Values of $\alpha_k$ and $\nu_k$ per transformation}} & $\forall_k$ & $\forall_k$ & $\forall_k$ & $\forall_k$ &  \multicolumn{2}{c:}{$\forall_k$}&  $\forall_k$ & $\forall_k$ &   \multicolumn{3}{c}{}\\
 
 \multicolumn{1}{c||}{}  & \multicolumn{2}{c|}{$\Psi_k=I$} & \multicolumn{2}{c|}{$\Psi_k=exp$} & \multicolumn{2}{c||}{$\Psi_k=I^*$}&&&&&&&&&&\\
 \hline
	$k$	&	$\alpha_k$	&	$\eta_k$	&	$\alpha_k$	&	$\eta_k$	&	$\alpha_k$	&	$\eta_k$	&		&	$\beta_k$	&	$\beta_k$	&	$\nu_k$	&	$\Psi_{k_0}$	&	$\Psi_{k_1}$	&	$\beta_k$	&	$\nu_k$	&	$\Psi_k$	&	$\beta_k$	&	$\nu_k$	\\
\hline
1	&	617.8	&	509.7	&	6.19	&	0.65	&	604.4	&	439.2	&		&		&		&		&	exp	&	exp	&		&		&	exp	&		&		\\
2	&	276.9	&	296.3	&	5.33	&	0.87	&	301.1	&	322.4	&		&		&		&		&	exp	&	exp	&		&		&	exp	&		&		\\
3	&	2.61	&	14.94	&	-1.86	&	1.53	&	0.50	&	1.55	&		&		&	$\sigma_3$	&		&	exp	&	exp	&		&		&	$I$	&		&		\\
4	&	6.94	&	4.81	&	1.62	&	0.95	&	7.90	&	9.52	&		&		&		&	-0.15	&	$I^*$	&	exp	&	-0.29	&	-0.15	&	exp	&	-0.29	&	-0.15	\\
5	&	72.08	&	16.72	&	4.25	&	0.23	&	72.13	&	17.02	&		&		&		&	-0.25	&	exp	&	exp	&		&	-0.25	&	$I$	&		&	-0.25	\\
6	&	16.69	&	17.28	&	2.27	&	1.21	&	20.23	&	36.99	&		&	$\sigma_6$	&	$\sigma_6$	&		&	$I^*$	&	exp	&		&		&	exp	&		&		\\
7	&	3.25	&	1.28	&	1.11	&	0.38	&	3.27	&	1.30	&		&		&		&	0.15	&	exp	&	exp	&	-0.44	&	0.15	&	exp	&	-0.44	&	0.15	\\
8	&	5.94	&	2.73	&	1.69	&	0.42	&	5.94	&	2.63	&		&		&		&		&	$I^*$	&	exp	&		&		&	exp	&		&		\\
9	&	11.66	&	13.59	&	1.84	&	1.22	&	13.29	&	24.78	&		&		&	$\sigma_9$	&		&	exp	&	exp	&	-0.41	&		&	exp	&	-0.41	&		\\
10	&	1.41	&	0.38	&	0.31	&	0.26	&	1.42	&	0.38	&		&		&		&		&	exp	&	exp	&	-0.14	&		&	exp	&	-0.14	&		\\
11	&	62.29	&	20.64	&	4.07	&	0.37	&	62.73	&	23.78	&		&		&		&		&	exp	&	exp	&		&		&	$I$	&		&		\\
12	&	592.1	&	1395	&	5.90	&	0.86	&	526.6	&	549.8	&		&		&		&		&	exp	&	exp	&		&		&	exp	&		&		\\
13	&	103.1	&	129.9	&	3.88	&	1.36	&	121.7	&	279.9	&		&	$\sigma_{13}$	&	$\sigma_{13}$	&	0.15	&	$I^*$	&	exp	&		&	0.15	&	exp	&		&	0.15	\\
14	&	177.4	&	61.28	&	5.13	&	0.31	&	177.0	&	55.50	&		&		&		&		&	$I^*$	&	exp	&		&		&	exp	&		&		\\
15	&	53.88	&	29.79	&	3.87	&	0.47	&	53.74	&	26.80	&		&		&		&	-0.15	&	exp	&	exp	&		&	-0.15	&	exp	&		&	-0.15	\\
16	&	8.55	&	0.76	&	2.14	&	0.09	&	8.56	&	0.78	&		&		&		&	0.10	&	exp	&	exp	&		&	0.10	&	$I$	&		&	0.10	\\
17	&	12.97	&	11.29	&	2.30	&	0.69	&	12.62	&	9.84	&		&		&	$\sigma_{17}$	&		&	exp	&	exp	&		&		&	exp	&		&		\\
18	&	0.71	&	0.48	&	-0.47	&	0.51	&	0.71	&	0.39	&		&		&		&		&	exp	&	exp	&		&		&	exp	&		&		\\
19	&	0.37	&	1.78	&	1.47	&	0.78	&	5.93	&	5.45	&		&		&		&		&	exp	&	exp	&		&		&	$I$	&		&		\\
20	&	0.78	&	1.11	&	-1.54	&	2.01	&	1.63	&	12.27	&		&	$\sigma_{20}$	&	$\sigma_{20}$	&		&	exp	&	exp	&		&		&	exp	&		&		\\
21	&	33.24	&	19.59	&	3.37	&	0.51	&	33.17	&	18.05	&		&		&		&	0.20	&	exp	&	exp	&		&	0.20	&	$I$	&		&	0.20	\\
22	&	0.31	&	0.20	&	-1.30	&	0.58	&	0.32	&	0.21	&		&		&		&	-0.20	&	exp	&	exp	&		&	-0.20	&	$I$	&		&	-0.20	\\
23	&	0.34	&	0.23	&	-1.29	&	0.71	&	0.35	&	0.29	&		&		&	$\sigma_{23}$	&		&	$I^*$	&	exp	&		&		&	exp	&		&		\\
24	&	0.22	&	0.29	&	-1.87	&	0.80	&	0.21	&	0.20	&		&		&		&		&	exp	&	exp	&		&		&	exp	&		&		\\
25	&	0.07	&	0.10	&	-2.82	&	0.64	&	0.07	&	0.05	&		&		&		&		&	exp	&	exp	&		&		&	exp	&		&		\\
26	&	3.64	&	2.12	&	1.04	&	1.01	&	4.72	&	6.29	&		&		&		&		&	$I^*$	&	exp	&	0.32	&		&	exp	&	0.32	&		\\
27	&	66.95	&	82.64	&	3.37	&	1.82	&	153.1	&	794.2	&		&	$\sigma_{27}$	&	$\sigma_{27}$	&		&	exp	&	exp	&		&		&	exp	&		&		\\
28	&	4.98	&	2.34	&	1.39	&	0.92	&	6.10	&	7.02	&		&		&		&	0.10	&	exp	&	exp	&		&	0.10	&	exp	&		&	0.10	\\
29	&	21.40	&	29.97	&	2.64	&	0.81	&	19.47	&	18.87	&		&		&		&		&	exp	&	exp	&	0.26	&		&	exp	&	0.26	&		\\
30	&	13.09	&	24.77	&	1.71	&	1.36	&	14.03	&	32.74	&		&		&	$\sigma_{30}$	&		&	$I^*$	&	exp	&		&		&	exp	&		&		\\
31	&	14.69	&	12.06	&	2.39	&	0.82	&	15.23	&	14.84	&		&		&		&		&	exp	&	exp	&	0.31	&		&	exp	&	0.31	&		\\
32	&	7.28	&	5.65	&	1.77	&	0.65	&	7.25	&	5.26	&		&		&		&		&	exp	&	exp	&		&		&	exp	&		&		\\
33	&	15.37	&	37.54	&	1.67	&	1.38	&	13.69	&	32.66	&		&		&		&		&	exp	&	exp	&		&		&	exp	&		&		\\
34	&	0.13	&	0.20	&	-2.64	&	1.21	&	0.15	&	0.27	&		&	$\sigma_{34}$	&	$\sigma_{34}$	&		&	exp	&	exp	&		&		&	exp	&		&		\\
35	&	22.53	&	37.47	&	2.62	&	0.94	&	21.27	&	25.29	&		&		&		&		&	$I^*$	&	exp	&		&		&	exp	&		&		\\
\bottomrule
\multicolumn{18}{l}{} \\
\multicolumn{7}{l}{\textbf{Nr. relevant biomarkers}} & 0 & 5 & 10 & 9 &\multicolumn{2}{c:}{9} & 7 & 9 & \multicolumn{3}{:c}{14} \\
\multicolumn{7}{l}{} &&&&&&&&&&\\
\multicolumn{7}{l}{\textbf{Nr. of participants (n)}} & 100 & 100 & 100 & 100 &\multicolumn{2}{c:}{100} & a: 100 & a: 100 & \multicolumn{3}{c}{a: 100} \\
\multicolumn{7}{l}{} &  &  &  &  &\multicolumn{2}{c:}{} & b: 400 & b: 400 & \multicolumn{3}{c}{b: 400} \\
\multicolumn{7}{l}{} &  &  &  &  &\multicolumn{2}{c:}{} & c: 250 & c: 250 & \multicolumn{3}{c}{c: 250} \\
\multicolumn{7}{l}{} &&&&&&&&&&\\
\multicolumn{7}{l}{\textbf{Proportion of cases ($\phi$)}} & 1/2 & 1/2 & 1/2 & 1/2 &\multicolumn{2}{c:}{1/2} & a: 1/2 & a: 1/2 & \multicolumn{3}{c}{a: 1/2} \\
\multicolumn{7}{l}{} &  &  &  &  &\multicolumn{2}{c:}{} & b: 1/2 & b: 1/2 & \multicolumn{3}{c}{b: 1/2} \\
\multicolumn{7}{l}{} &  &  &  &  &\multicolumn{2}{c:}{} & c: 1/5 & c: 1/5 & \multicolumn{3}{c}{c: 1/5} \\
\multicolumn{7}{l}{} &&&&&&&&&&\\
\bottomrule
    \end{tabular}}
\end{table}%

\subsection{Results}
\begin{figure}[H]
 \centering
  \includegraphics[width=1\linewidth]{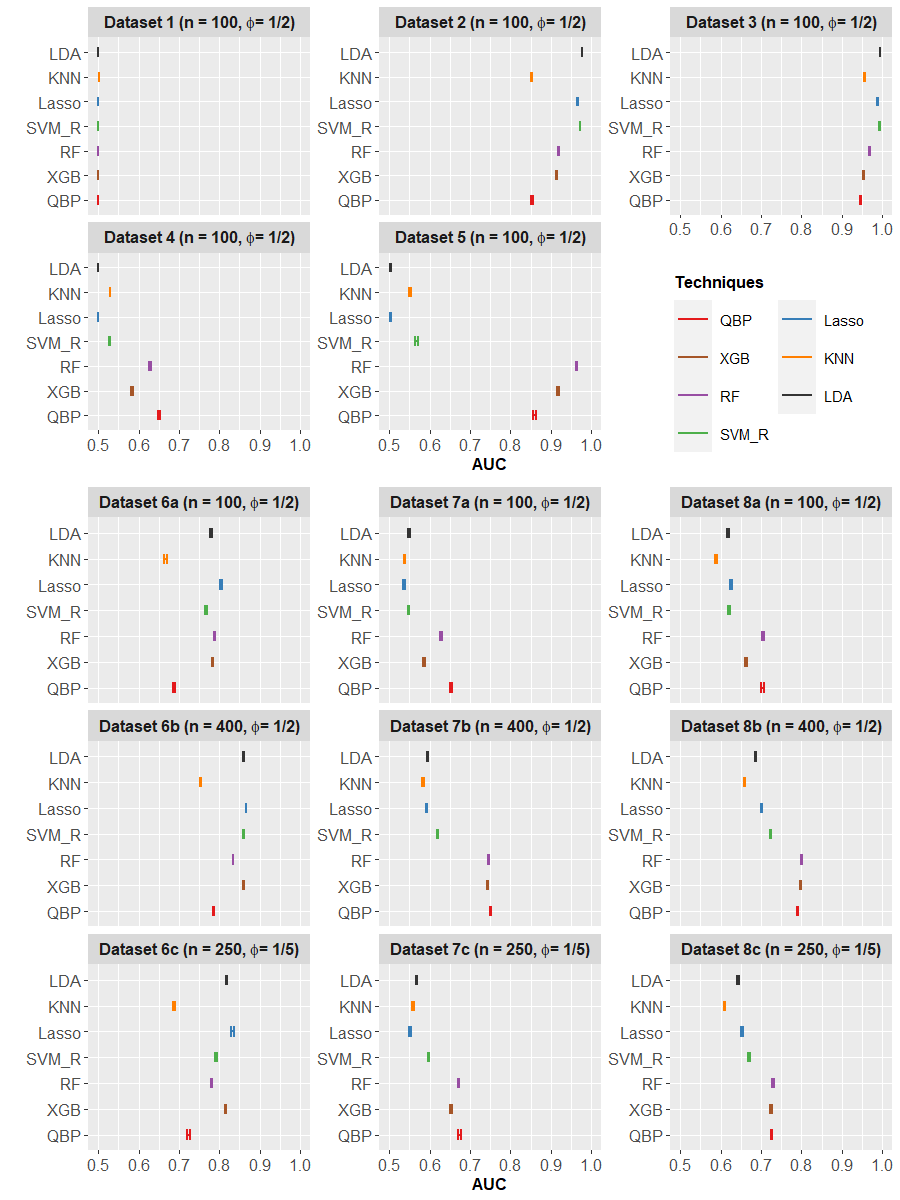}
\caption{Confidence intervals of the classification performances (in AUC) based on 500 simulations per dataset type}
\label{fig:VALIDATION_PERF_SIMULATION_CI}
\end{figure}

For all binary classification techniques and datasets, the predictive performance is presented in Table~\ref{tab:SimulatedPerformanceVAL}. In Figure~\ref{fig:Density_AUC.VAL_DEPR_DS} and \ref{fig:VALIDATION_PERF_SIMULATION_CI}, we present the density plots and confidence intervals of the predictive performance, respectively. These graphs only contain a subset of the techniques, namely PLR.Lasso, LDA, SVM.Radial, RF, kNN, XGB and QBP. This selection is based on superior performance in at least one of the simulated datasets.

Moreover, in Table~\ref{tab:NumberOfComponentsSimulation} the number of used biomarkers in the final model is presented as well as the applied number of sparse components for the PCLR and PLS-LDA and the number of neighbors for kNN. The effect of a sample size on the sensitivity, specificity and accuracy of the biomarker selection is presented in Table~\ref{tab:EffectSampleSize}. Here, only the methods PLR.Lasso, PLR.EN and QBP are included in the overview, since all the other methods always include all biomarkers and therefore apply no selection. Finally, in Table~\ref{tab:computation times} the average computation times are listed for the datasets with a sample size of $n=100$, $n=250$ and $n=400$.

\captionsetup[subfigure]{font=scriptsize,labelfont=scriptsize}

\begin{table}[H]
	\small
  \centering
 \caption{Performance (in AUC) of all considered techniques of all simulated datasets on validation data}
 \scalebox{0.75}{
    \begin{tabular}{llccccccccccccc}
        \toprule
 \multicolumn{2}{l}{\textbf{Validation data}}  && \multicolumn{3}{c}{PLR} &&\multirow{2}[0]{*}{PLS-}&& \multicolumn{2}{c}{SVM} &&&&\\
    \cmidrule(r){4-6} \cmidrule(r){10-11}
Dataset	&													&	LR		&	Lasso	&	EN		&	Ridge	&	PCLR	&LDA	&	LDA		&	Linear	&	Radial	&	RF		&	kNN		&	XGB		&	QBP	\\
\midrule
\multirow{2}[0]{*}{1 ($n=100, \; \phi=1/2$)}	&	mean	&	0.500	&	0.500	&	0.500	&	0.500	&	0.500	&	0.500	&	0.500	&	0.500	&	0.500	&	0.500	&	0.500	&	0.500	&	0.500   \\
												&	sd		&	0.008	&	0.008	&	0.008	&	0.008	&	0.008	&	0.008	&	0.008	&	0.008	&	0.008	&	0.008	&	0.008	&	0.008	&	0.008	\\
		\rule{0pt}{3ex} 
\multirow{2}[0]{*}{2 ($n=100, \; \phi=1/2$)}	&	mean	&	0.943	&	0.967	&	0.971	&	0.972	&	0.937	&	0.977	&	0.977	&	0.971	&	0.973	&	0.919	&	0.853	&	0.915	&	0.854	\\
												&	sd		&	0.022	&	0.013	&	0.011	&	0.011	&	0.029	&	0.009	&	0.008	&	0.010	&	0.010	&	0.015	&	0.026	&	0.014	&	0.034	\\
		\rule{0pt}{3ex} 
\multirow{2}[0]{*}{3 ($n=100, \; \phi=1/2$)}	&	mean	&	0.984	&	0.988	&	0.992	&	0.992	&	0.980	&	0.995	&	0.996	&	0.994	&	0.993	&	0.968	&	0.957	&	0.953	&	0.948	\\
												&	sd		&	0.009	&	0.007	&	0.006	&	0.005	&	0.013	&	0.003	&	0.002	&	0.004	&	0.005	&	0.008	&	0.013	&	0.01	&	0.017	\\
		\rule{0pt}{3ex} 
\multirow{2}[0]{*}{4 ($n=100, \; \phi=1/2$)}	&	mean	&	0.499	&	0.499	&	0.499	&	0.499	&	0.500	&	0.499	&	0.499	&	0.500	&	0.527	&	0.629	&	0.530	&	0.584	&	0.652	\\
												&	sd		&	0.008	&	0.008	&	0.008	&	0.008	&	0.008	&	0.008	&	0.008	&	0.008	&	0.017	&	0.029	&	0.011	&	0.031	&	0.028	\\
		\rule{0pt}{3ex} 
\multirow{2}[0]{*}{5 ($n=100, \; \phi=1/2$)}	&	mean	&	0.502	&	0.504	&	0.505	&	0.502	&	0.501	&	0.502	&	0.503	&	0.503	&	0.568	&	0.963	&	0.552	&	0.917	&	0.860	\\
												&	sd		&	0.009	&	0.014	&	0.016	&	0.009	&	0.009	&	0.009	&	0.010	&	0.010	&	0.039	&	0.013	&	0.018	&	0.027	&	0.035	\\
		\rule{0pt}{3ex} 
\multirow{2}[0]{*}{6a ($n=100, \; \phi=1/2$)}	&	mean	&	0.714	&	0.804	&	0.801	&	0.782	&	0.730	&	0.776	&	0.779	&	0.763	&	0.768	&	0.788	&	0.667	&	0.783	&	0.688	\\
												&	sd		&	0.039	&	0.031	&	0.027	&	0.031	&	0.043	&	0.033	&	0.029	&	0.033	&	0.034	&	0.021	&	0.037	&	0.023	&	0.036	\\
		\rule{0pt}{3ex} 
\multirow{2}[0]{*}{6b ($n=400, \; \phi=1/2$)}	&	mean	&	0.862	&	0.866	&	0.865	&	0.861	&	0.859	&	0.859	&	0.860	&	0.854	&	0.859	&	0.834	&	0.754	&	0.859	&	0.785	\\
												&	sd		&	0.009	&	0.009	&	0.009	&	0.009	&	0.011	&	0.010	&	0.010	&	0.011	&	0.011	&	0.010	&	0.018	&	0.010	&	0.015	\\
		\rule{0pt}{3ex} 
\multirow{2}[0]{*}{6c ($n=250, \; \phi=1/5$)}	&	mean	&	0.824	&	0.832	&	0.831	&	0.826	&	0.816	&	0.815	&	0.818	&	0.780	&	0.792	&	0.781	&	0.688	&	0.815	&	0.724	\\
												&	sd		&	0.020	&	0.037	&	0.020	&	0.019	&	0.024	&	0.021	&	0.019	&	0.029	&	0.027	&	0.020	&	0.030	&	0.016	&	0.038	\\
		\rule{0pt}{3ex} 
\multirow{2}[0]{*}{7a ($n=100, \; \phi=1/2$)}	&	mean	&	0.551	&	0.537	&	0.538	&	0.545	&	0.536	&	0.544	&	0.549	&	0.542	&	0.547	&	0.629	&	0.539	&	0.587	&	0.652	\\
												&	sd		&	0.023	&	0.027	&	0.027	&	0.020	&	0.024	&	0.021	&	0.022	&	0.023	&	0.019	&	0.029	&	0.017	&	0.029	&	0.029	\\
		\rule{0pt}{3ex} 
\multirow{2}[0]{*}{7b ($n=400, \; \phi=1/2$)}	&	mean	&	0.601	&	0.592	&	0.592	&	0.594	&	0.588	&	0.590	&	0.596	&	0.582	&	0.620	&	0.746	&	0.584	&	0.744	&	0.751	\\
												&	sd		&	0.014	&	0.023	&	0.023	&	0.017	&	0.026	&	0.018	&	0.014	&	0.015	&	0.018	&	0.015	&	0.018	&	0.016	&	0.021	\\
		\rule{0pt}{3ex} 
\multirow{2}[0]{*}{7c ($n=250, \; \phi=1/5$)}	&	mean	&	0.574	&	0.552	&	0.552	&	0.561	&	0.548	&	0.560	&	0.568	&	0.563	&	0.598	&	0.672	&	0.559	&	0.653	&	0.674	\\
												&	sd		&	0.021	&	0.033	&	0.033	&	0.024	&	0.032	&	0.024	&	0.022	&	0.023	&	0.021	&	0.020	&	0.020	&	0.030	&	0.029	\\
		\rule{0pt}{3ex} 
\multirow{2}[0]{*}{8a ($n=100, \; \phi=1/2$)}	&	mean	&	0.623	&	0.626	&	0.627	&	0.629	&	0.618	&	0.623	&	0.618	&	0.603	&	0.621	&	0.705	&	0.588	&	0.663	&	0.704	\\
												&	sd		&	0.032	&	0.038	&	0.036	&	0.030	&	0.038	&	0.030	&	0.031	&	0.032	&	0.03	&	0.028	&	0.026	&	0.029	&	0.029	\\
		\rule{0pt}{3ex} 
\multirow{2}[0]{*}{8b ($n=400, \; \phi=1/2$)}	&	mean	&	0.703	&	0.702	&	0.700	&	0.698	&	0.698	&	0.687	&	0.686	&	0.668	&	0.723	&	0.799	&	0.659	&	0.798	&	0.791	\\
												&	sd		&	0.015	&	0.019	&	0.017	&	0.016	&	0.017	&	0.017	&	0.016	&	0.018	&	0.018	&	0.013	&	0.020	&	0.014	&	0.016	\\
		\rule{0pt}{3ex} 
\multirow{2}[0]{*}{8c ($n=250, \; \phi=1/5$)}	&	mean	&	0.663	&	0.652	&	0.651	&	0.650	&	0.651	&	0.641	&	0.643	&	0.628	&	0.670	&	0.730	&	0.610	&	0.725	&	0.726	\\
												&	sd		&	0.024	&	0.034	&	0.035	&	0.026	&	0.033	&	0.026	&	0.024	&	0.029	&	0.024	&	0.020	&	0.025	&	0.024	&	0.026	\\
\bottomrule
 \end{tabular}}
\label{tab:SimulatedPerformanceVAL}%
\end{table}%

\begin{figure}[H]
 \centering
  \includegraphics[width=1\linewidth]{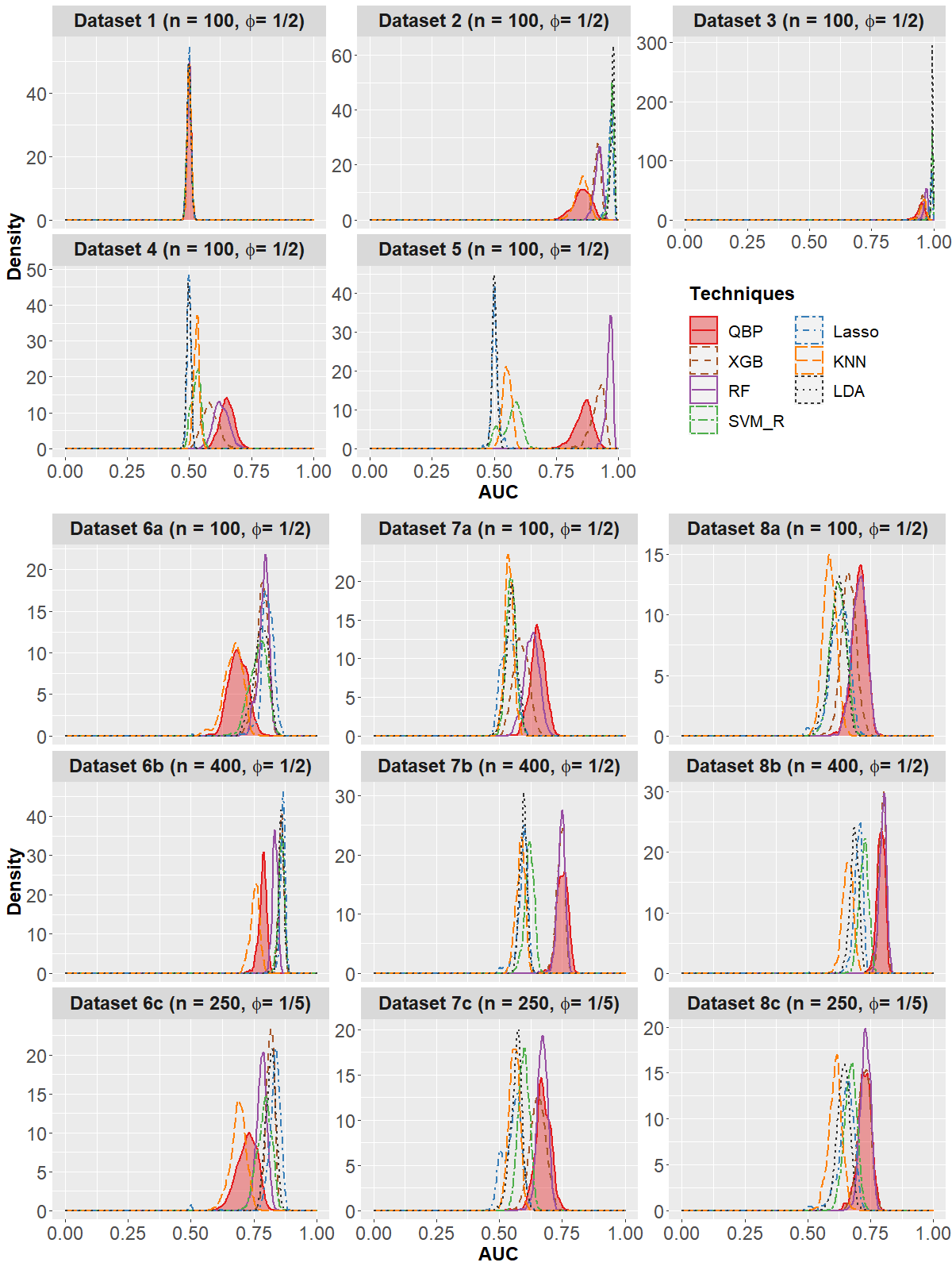}
\caption{Performance (in AUC) of all 500 simulations with each a different training set to tune the parameters and validation set of 5000 subjects to assess the performance}
\label{fig:VALIDATION_PERF_SIMULATION}
\end{figure}

\begin{table}[htbp]
\small
  \centering
  \caption{Number of included biomarkers, components for sparse methods derived from training data and applied in final model and neighbors included for kNN. \textit{Note that LR.LOGIT, PLR.Ridge, PCLR, PLS-LDA, LDA, SVM, RF, kNN $\&$ XGB use all biomarkers: $mean=35$ and $sd=0$.}}
  \scalebox{0.75}{
    \begin{tabular}{lclcccccc}
    \toprule
    &&  & \multicolumn{3}{c}{\textbf{Biomarkers}} & \multicolumn{2}{c}{\textbf{Sparse components}} &\\
    \cmidrule(r){4-6} \cmidrule(r){7-8}
             & Nr. relevant && \multicolumn{2}{c}{PLR}& && &\\
    \cmidrule(r){4-5}
Dataset	&	biomarkers	&		&	Lasso	&	EN	&	QBP	&	PCLR 	&	PLS-LDA	&	kNN	\\
    \midrule
\multirow{2}[0]{*}{1 ($n=100, \; \phi=1/2$)}	&	\multirow{2}[0]{*}{0}	&	mean	&	16	&	16.5	&	26	&	15	&	5.4	&	12	\\
	&		&	sd	&	13.6	&	13.8	&	6.9	&	11.5	&	5.6	&	9.2	\\
			\rule{0pt}{3ex} 
\multirow{2}[0]{*}{2 ($n=100, \; \phi=1/2$)}	&	\multirow{2}[0]{*}{5}	&	mean	&	18.5	&	23	&	21.8	&	29.1	&	6.1	&	19.8	\\
	&		&	sd	&	5.9	&	7.4	&	5	&	4.6	&	3.0	&	6.6	\\
			\rule{0pt}{3ex} 
\multirow{2}[0]{*}{3 ($n=100, \; \phi=1/2$)}	&	\multirow{2}[0]{*}{10}	&	mean	&	18.8	&	23.5	&	25	&	26.3	&	4.7	&	19.4	\\
	&		&	sd	&	4.5	&	6.4	&	4.6	&	6.4	&	3.1	&	6.7	\\
			\rule{0pt}{3ex} 
\multirow{2}[0]{*}{4 ($n=100, \; \phi=1/2$)}	&	\multirow{2}[0]{*}{9}	&	mean	&	16.4	&	16.8	&	27.2	&	14.6	&	5.4	&	10.4	\\
	&		&	sd	&	13.6	&	13.9	&	6	&	11.5	&	5.3	&	8.4	\\
			\rule{0pt}{3ex} 
\multirow{2}[0]{*}{5 ($n=100, \; \phi=1/2$)}	&	\multirow{2}[0]{*}{9}	&	mean	&	14.2	&	14.8	&	22.9	&	16	&	4.8	&	9.7	\\
	&		&	sd	&	13.4	&	13.6	&	4.8	&	11.6	&	4.6	&	8.2	\\
			\rule{0pt}{3ex} 
\multirow{2}[0]{*}{6a ($n=100, \; \phi=1/2$)}	&	\multirow{2}[0]{*}{7}	&	mean	&	11.7	&	12.5	&	24.2	&	24.4	&	5.4	&	18.9	\\
	&		&	sd	&	9.6	&	10.9	&	6.3	&	6.8	&	3.8	&	7.7	\\
			\rule{0pt}{3ex} 
\multirow{2}[0]{*}{6b ($n=400, \; \phi=1/2$)}	&	\multirow{2}[0]{*}{7}	&	mean	&	24.3	&	26.2	&	6.8	&	33.6	&	6.4	&	47.7	\\
	&		&	sd	&	5.6	&	5.7	&	4.2	&	2.1	&	2.9	&	9.6	\\
			\rule{0pt}{3ex} 
\multirow{2}[0]{*}{6c ($n=250, \; \phi=1/5$)}	&	\multirow{2}[0]{*}{7}	&	mean	&	18.5	&	20.4	&	15.7	&	31.8	&	6.0	&	32.8	\\
	&		&	sd	&	9.2	&	10.3	&	7.3	&	3.1	&	3.0	&	10.9	\\
			\rule{0pt}{3ex} 
\multirow{2}[0]{*}{7a ($n=100, \; \phi=1/2$)}	&	\multirow{2}[0]{*}{9}	&	mean	&	20.3	&	20.7	&	27.4	&	20.2	&	5.8	&	11.8	\\
	&		&	sd	&	13.4	&	13.6	&	5.8	&	11.7	&	5.0	&	8.6	\\
			\rule{0pt}{3ex} 
\multirow{2}[0]{*}{7b ($n=400, \; \phi=1/2$)}	&	\multirow{2}[0]{*}{9}	&	mean	&	28.1	&	28.6	&	16.9	&	30.2	&	7.2	&	21.9	\\
	&		&	sd	&	9.6	&	9.8	&	4.4	&	9	&	3.9	&	15.1	\\
			\rule{0pt}{3ex} 
\multirow{2}[0]{*}{7c ($n=250, \; \phi=1/5$)}	&	\multirow{2}[0]{*}{9}	&	mean	&	21.4	&	22.3	&	23.2	&	21.1	&	5.4	&	19	\\
	&		&	sd	&	13.2	&	13.3	&	7.5	&	12.9	&	4.1	&	13.5	\\
			\rule{0pt}{3ex} 
\multirow{2}[0]{*}{8a ($n=100, \; \phi=1/2$)}	&	\multirow{2}[0]{*}{14}	&	mean	&	17.5	&	18.4	&	28.7	&	21.3	&	3.9	&	13.4	\\
	&		&	sd	&	12.3	&	12.5	&	5.3	&	9.1	&	3.6	&	8.5	\\
			\rule{0pt}{3ex} 
\multirow{2}[0]{*}{8b ($n=400, \; \phi=1/2$)}	&	\multirow{2}[0]{*}{14}	&	mean	&	20.8	&	21.2	&	19.7	&	29.9	&	3.7	&	34	\\
	&		&	sd	&	10.2	&	10.7	&	3.9	&	4.9	&	2.6	&	14.9	\\
			\rule{0pt}{3ex} 
\multirow{2}[0]{*}{8c ($n=250, \; \phi=1/5$)}	&	\multirow{2}[0]{*}{14}	&	mean	&	21.9	&	22.6	&	24.4	&	27	&	3.6	&	22.7	\\
	&		&	sd	&	11.9	&	12.2	&	6.8	&	7.4	&	2.7	&	13.2	\\

 	\bottomrule
    \end{tabular}}
  \label{tab:NumberOfComponentsSimulation}%
\end{table}%

\begin{table}[H]
  \small
  \centering
  \caption{Effect of sample size on inclusion performance of the relevant biomarkers. \textit{Note that the methods LR.LOGIT, PLR.Ridge, PCLR, PLS-LDA, LDA, SVM, RF, kNN $\&$ XGB use all biomarkers and are not included in this overview.}}
   \scalebox{0.75}{ 
    \begin{tabular}{cclccc}
    \toprule
                  &Nr. relevant&& \multicolumn{2}{c}{PLR} & \\
                  \cmidrule{4-5}  
Dataset	& biomarkers	&	measure	&	Lasso	&	EN	&	QBP	\\
\midrule
\multirow{3}[0]{*}{6a ($n=100, \; \phi=1/2$)}	&	\multirow{3}[0]{*}{7}	&	accuracy	&	0.689	&	0.676	&	0.459	\\
	&		&	sensitivity	&	0.557	&	0.58	&	0.878	\\
	&		&	specificity	&	0.721	&	0.7	&	0.355	\\
			\rule{0pt}{3ex} 
\multirow{3}[0]{*}{6b ($n=400, \; \phi=1/2$)}	&	\multirow{3}[0]{*}{7}	&	accuracy	&	0.467	&	0.42	&	0.885	\\
	&		&	sensitivity	&	0.903	&	0.922	&	0.696	\\
	&		&	specificity	&	0.358	&	0.295	&	0.932	\\
			\rule{0pt}{3ex} 
\multirow{3}[0]{*}{6c ($n=250, \; \phi=1/5$)}	&	\multirow{3}[0]{*}{7}	&	accuracy	&	0.58	&	0.537	&	0.666	\\
	&		&	sensitivity	&	0.771	&	0.797	&	0.791	\\
	&		&	specificity	&	0.533	&	0.472	&	0.635	\\
			\rule{0pt}{3ex} 
\multirow{3}[0]{*}{7a ($n=100, \; \phi=1/2$)}	&	\multirow{3}[0]{*}{9}	&	accuracy	&	0.481	&	0.475	&	0.434	\\
	&		&	sensitivity	&	0.618	&	0.628	&	0.922	\\
	&		&	specificity	&	0.434	&	0.422	&	0.266	\\
			\rule{0pt}{3ex} 
\multirow{3}[0]{*}{7b ($n=400, \; \phi=1/2$)}	&	\multirow{3}[0]{*}{9}	&	accuracy	&	0.395	&	0.383	&	0.744	\\
	&		&	sensitivity	&	0.883	&	0.887	&	0.942	\\
	&		&	specificity	&	0.226	&	0.208	&	0.676	\\
			\rule{0pt}{3ex} 
\multirow{3}[0]{*}{7c ($n=250, \; \phi=1/5$)}	&	\multirow{3}[0]{*}{9}	&	accuracy	&	0.482	&	0.465	&	0.545	\\
	&		&	sensitivity	&	0.68	&	0.698	&	0.903	\\
	&		&	specificity	&	0.414	&	0.385	&	0.421	\\
			\rule{0pt}{3ex} 
\multirow{3}[0]{*}{8a ($n=100, \; \phi=1/2$)}	&	\multirow{3}[0]{*}{14}	&	accuracy	&	0.556	&	0.552	&	0.513	\\
	&		&	sensitivity	&	0.569	&	0.599	&	0.917	\\
	&		&	specificity	&	0.548	&	0.522	&	0.244	\\
			\rule{0pt}{3ex} 
\multirow{3}[0]{*}{8b ($n=400, \; \phi=1/2$)}	&	\multirow{3}[0]{*}{14}	&	accuracy	&	0.59	&	0.586	&	0.787	\\
	&		&	sensitivity	&	0.732	&	0.738	&	0.939	\\
	&		&	specificity	&	0.495	&	0.485	&	0.687	\\
			\rule{0pt}{3ex} 
\multirow{3}[0]{*}{8c ($n=250, \; \phi=1/5$)}	&	\multirow{3}[0]{*}{14}	&	accuracy	&	0.54	&	0.534	&	0.619	\\
	&		&	sensitivity	&	0.709	&	0.725	&	0.894	\\
	&		&	specificity	&	0.427	&	0.406	&	0.435	\\

          \bottomrule
    \end{tabular}}
    \label{tab:EffectSampleSize}%
\end{table}%

\begin{table}[H]
\small
\centering
\caption{Average computation times (in seconds) for a training (6-fold inner cross-validation) and validation (model fit and evaluation) cycle for a single datasets with a sample size of $n=100$ (dataset 1,2,3,4,5,6a,7a,8a), $n=400$ (dataset 6b,7b,8b) and $n=250$ (dataset 6c,7c,8c). \textit{Note that LR.LOGIT and LDA do not include a tunable parameter optimization.}}
\label{tab:computation times}
\scalebox{0.75}{ 
\begin{tabular}{lccccccccccccc}
\toprule
              &  &        \multicolumn{3}{c}{PLR}  &&&& \multicolumn{2}{c}{SVM} &&&&\\
    \cmidrule(r){3-5} \cmidrule(r){9-10}
Sample size	&	LR		&	Lasso	&	EN		&	Ridge	&	PCLR	&	PLS-LDA	&	LDA		&	Linear	&	Radial	&	RF		&	kNN		&	XGB		&	QBP\\
\midrule
100			&	0.06	&	0.81	&	0.46	&	0.98	&	9.09	&	8.03	&	0.01	&	81.1	&	5.64	&	18.7	&	94.5	&	80.5	&	22.6\\
400			&	0.06	&	0.39	&	0.40	&	0.67	&	9.71	&	9.64	&	0.02	&	363.9	&	31.6	&	94.9	&	135.6 	&	86.5	&	25.4\\
250			&	0.06	&	0.48	&	0.43	&	0.70	&	9.51	&	8.89	&	0.01	&	367.2	&	14.0	&	51.4	&	125.8	&	82.4	&	24.3\\
 \hline
\end{tabular}}
\end{table}

\section{Case study}

\subsection{Major Depression Disorder}
\subsubsection{Design of the study}

The MDD data contains 35 biomarkers, of which 16 are serum based biomarkers and 19 are urine based biomarkers. An overview of all biomarker types is presented in Table~\ref{tab:IncludedbiomarkersMDD}. These serum and first morning urine biomarkers were selected based on a thorough literature search, combined with a pilot study in 24 participants (12 MDD patients and their sex, age and ethnic matched non-MDD controls). The MDD study contains 101 patients in total, of which 4 patients had missing values. These patients were excluded from the analysis to make a fair comparison between the methods and avoid the effect of imputations on the performance. The predictive performance of all methods is assessed using rdCV.

\begin{table}[H]
	\small
  \centering
  \caption{Included biomarkers in MDD data, where the numbers are aligned with the biomarker numbers $k$.}
  \scalebox{0.87}{
    \begin{tabular}{ll|lll}
    \toprule
    \multicolumn{2}{c}{\textbf{Serum biomarkers}} & \multicolumn{3}{c}{\textbf{Urine biomarker}} \\
    \midrule
1. BDNF	&	9. Thromboxane	&	17. cAMP	&	25. Endothelin	&	33. Lipocalin	\\
2. Midkine	&	10. Endothelin	&	18. cGMP	&	26. Aldosteron	&	34. Pregnonelon	\\
3. Nitrotyrosine	&	11. Lipocalin	&	19. Calprotectin	&	27. Adiponectin	&	35. NPY	\\
4. EGF	&	12. NPY	&	20. Leptin	&	28. HVEM	&				\\
5. TNFR2	&	13. Leptin	&	21. LTB4	&	29. Midkine	&				\\
6. LTB4	&	14. HVEM	&	22. Cortisol	&	30. EGF	&				\\
7. Cortisol	&	15. Vit-D	&	23. Thromboxane	&	31. SubstanceP	&				\\
8. Calprotectin	&	16. Zonulin	&	24. Isoprostane	&	32. TNFR2	&				\\
    \bottomrule
    \end{tabular}}
  \label{tab:IncludedbiomarkersMDD}%
\end{table}%

\subsubsection{Results}
\begin{figure}[H]
 \centering
 \includegraphics[width=.5\linewidth]{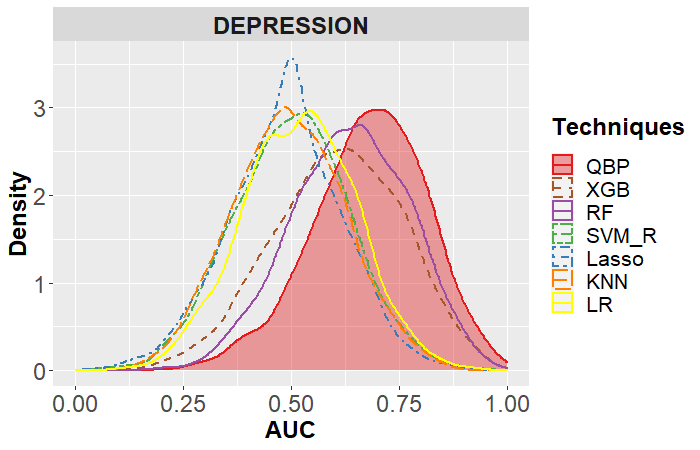}
 \caption{AUC of validation data}
  \label{fig:Density_AUC.VAL_DEPR.MMD}
 \caption{Performance (in AUC) measured on validation dataset: \\
 500 repeats, 6-fold outer CV, 6-fold inner CV}
\end{figure}

For all binary classification techniques, the predictive performance expressed in the mean and its standard error are shown in Table~\ref{tab:MDDperformance}. Besides, this table contains the number of biomarkers that were used on the validation dataset. The density plots of the predictive performance of a subset of the techniques (LR, PLR.Lasso, SVM.Radial, RF, kNN, XGB and QBP) are shown in \ref{fig:Density_AUC.VAL_DEPR.MMD}.

\begin{table}[H]
	\small
  \centering
  \caption{Summary statistics of the performance of all methods on MDD data: \\AUC validation data and included number of components \\
  \small{ $^*$ \textit{For PCLR, PLS-LDA the number of sparse components is given by $(\ldots)$ and for kNN the number of neighbors $k$ is represented by $(\ldots)$ }}}
    \scalebox{0.75}{
\begin{tabular}{llccccccccccccc}
\toprule
              &&  &        \multicolumn{3}{c}{PLR}  &&&& \multicolumn{2}{c}{SVM} &&&&\\
    \cmidrule(r){4-6} \cmidrule(r){10-11}
	&		&	LR	&	Lasso	&	EN	&	Ridge	&	PCLR	&	PLS-LDA	&	LDA	&	Linear	&	Radial	&	RF	&	kNN	&	XGB	&	QBP	\\
	\midrule
\multirow{2}[0]{*}{AUC VAL}	&	mean	&	0.518	&	0.486	&	0.485	&	0.512	&	0.492	&	0.505	&	0.523	&	0.516	&	0.501	&	0.635	&	0.495	&	0.606	&	0.680	\\
	&	sd	&	0.130	&	0.131	&	0.130	&	0.135	&	0.131	&	0.0.131	&	0.134	&	0.133	&	0.134	&	0.136	&	0.131	&	0.150	&	0.132	\\
\multirow{2}[0]{*}{NCOMP}	&	mean	&	35	&	22.2	&	22.7	&	35	&	35 (19.3)	&	35 (5.5)	&	35	&	35	&	35	&	35	&	35 (9.3)	&	35	&	27.6	\\
	&	sd	&	0	&	13.2	&	13.2	&	0	&	0 (10.9)	&	0 (5.4)	&	0	&	0	&	0	&	0	&	0 (6.6)	&	0	&	4.5	\\

          \bottomrule
    \end{tabular}}
\label{tab:MDDperformance}%
\end{table}%

\subsection{Trisomy}
\subsubsection{Design of the study}
The trisomy dataset is provided by the Foundation of Prenatal Screening of the Northern Netherlands and consists of a first-trimester combined-test screening program in the Netherlands in a multi-centre routine clinical setting. Whereas earlier evaluations have taken place based on data in the period of July 2002 to May 2004, as published in \cite{schielen2006multi}, this study only includes subjects after of July 1, 2010. From this moment, risks at trisomy were calculated by the Dutch National Institute for Public Health and the Environment (RIVM) according to the Astraia/Fetal Medicine Foundation (FMF) risk software. 

The first-trimester combined test is composed of three elements: (1) assay of the serum concentrations of pregnancy-associated plasma protein A ($PAPP\text{-}A$) and the free $\beta$ subunit of human chorion gonadotrophin ($f \beta\text{-}hCG$) between 8–14 weeks of the pregnancy, (2) ultrasound measurement of the nuchal translucency ($NT$) subcutaneous oedema in the fetal neck, to be measured at a gestational age (GA) between 10–11 and 14 weeks, and (3) maternal age. Accompanied with this test, the crown-rump length ($CRL$) that was used to determine the GA was recorded, the age of the mother, parity and gravidity. 

In the late '90s, with the introduction of maternal serum biochemistry and ultrasound screening for chromosomal defects at different stages of pregnancy, it has become necessary to establish maternal and gestational age-specific risks for chromosomal defect \cite{nicolaides2003screening}. Since the GA affects the biochemical parameters ($PAPP\text{-}A$ and $f\beta\text{-}hCG$), we use the multiple of median (MoM) versions $PAPP\text{-}A$ and $f\beta\text{-}hCG$ in the analysis. 

The method that RIVM uses to determine the risk on trisomy per subject, namely the FMF risk, takes into account women's a priori risk, based on her maternal age and gestational age, and multiply this by a series likelihood ratios of $MoM\text{-}f\beta\text{-}hCG$, $MoM\text{-}PAPP\text{-}A$, $NT$. This likelihood ratio is obtained by dividing the percentage of cases by the percentage of controls with that measurement. The probability on having Down Syndrome is defined in terms of an odds-ratio \cite{shiefa2013first}. 

In the dataset provided by RIVM, the FMF risk is determined on a dataset with $n=3784$ observations (53 cases and 3731 controls) and derived using the biomarkers maternal age, $NT$, $MoM\text{-}f\beta\text{-}hCG$ and $MoM\text{-}PAPP\text{-}A$. Note that for some subjects in this dataset a single biomarker value is missing. For these missing values of a certain combination of subject and biomarker, QBP imputes a disease score of 0, making that the biomarker distribution remains unaffected. As the classification performance of the FMF risk was assessed by training and validating on the full dataset, we do the same for QBP.

For the comparison of QBP with the selected alternative methods we use a smaller dataset with only complete observations to make sure that the comparison is not influenced by any imputation procedure. This dataset has $n=3514$ observations (48 cases and 3466 controls) and utilizes the biomarkers maternal age, parity, gravidity, $MoM\text{-}f\beta\text{-}hCG$, $MoM\text{-}PAPP\text{-}A$, $NT$ and $CRL$. Here, the predictive performance is assessed using rdCV. Here, QBP uses the optimal tunable parameter setting of the maximal interval score and lower boundary on the exceedratio.

\subsubsection{Results}
\begin{figure}[H]
 \centering
  \includegraphics[width=.50\linewidth]{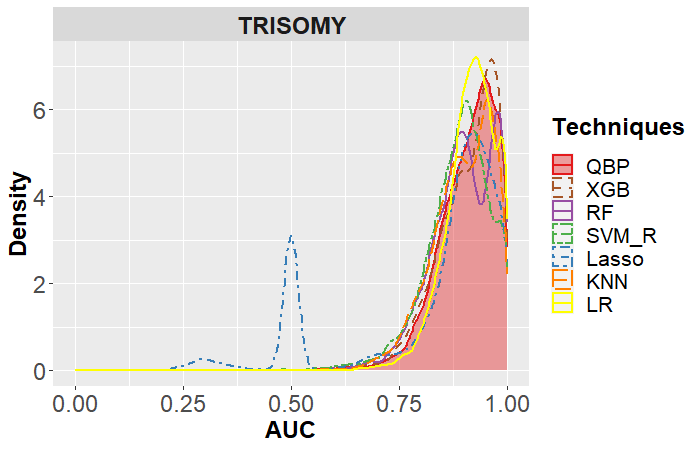}
 \caption{AUC of validation data}
  \label{fig:Density_AUC.VAL_DEPR_DS}
 \caption{Performance (in AUC) measured on validation dataset: \\
 500 repeats, 6-fold outer CV, 6-fold inner CV}
\end{figure}

The predictive performance and number of biomarkers of all considered techniques is presented in Table~\ref{tab:DOWNperformance}. In Figure~\ref{fig:Density_AUC.VAL_DEPR_DS}, the density plots of the predictive performance are provided for subset of the techniques - namely LR, PLR.Lasso, SVM.Radial, RF, kNN, XGB and QBP.

Regarding the FMF risk, we obtain a performance of the classification of cases and controls of $AUC=0.9151$. For QBP, we have $AUC=0.9249$ with the maximal interval score $v=(1,2,3)$ and lower boundaries for the exceedratios $R^*=(2,3,6)$ as optimal tunable parameter combination.

\begin{table}[H]
	\small
  \centering
  \caption{Summary statistics of the performance of all methods on Trisomy data: \\AUC validation data and included number of components \\
  \small{ $^*$ \textit{For PCLR, PLS-LDA the number of sparse components is given by $(\ldots)$ and for kNN the number of neighbors $k$ is represented by $(\ldots)$ }}}
    \scalebox{0.75}{
\begin{tabular}{llccccccccccccc}
\toprule
              &&  &        \multicolumn{3}{c}{PLR}  &&&& \multicolumn{2}{c}{SVM} &&&&\\
    \cmidrule(r){4-6} \cmidrule(r){10-11}
	&		&	LR	&	Lasso	&	EN	&	Ridge	&	PCLR	&	PLS-LDA	&	LDA	&	Linear	&	Radial	&	RF	&	kNN	&	XGB	&	QBP	\\
	\midrule
\multirow{2}[0]{*}{AUC VAL}	&	mean	&	0.914	&	0.834	&	0.855	&	0.728	&	0.909	&	0.881	&	0.886	&	0.909	&	0.886	&	0.898	&	0.896	&	0.908	&	0.908	\\
	&	sd	&	0.058	&	0.176	&	0.153	&	0.209	&	0.062	&	0.076	&	0.073	&	0.058	&	0.075	&	0.073	&	0.070	&	0.069	&	0.066	\\
\multirow{2}[0]{*}{NCOMP}	&	mean	&	7	&	5.0	&	6.4	&	7	&	7 (6.8)	&	7 (4.6)	&	7	&	7	&	7	&	7	&	7 (143.2)	&	7	&	5.8	\\
	&	sd	&	0	&	2.1	&	0.8	&	0	&	0 (0.7)	&	0 (2.0)	&	0	&	0	&	0	&	0	&	0 (24.8)	&	0	&	0.6	\\
      \bottomrule
    \end{tabular}}
\label{tab:DOWNperformance}%
\end{table}%

\section{Discussion}

In this study, we have performed an extensive comparative study between supervised binary disease prediction methods, focusing on all sorts of differences in distributions between cases and controls that appear in reality caused by biological processes and the complexity of diseases. Inspired by the situation in which using simple location measures are failing to discriminate between cases and controls, and using only tail information may better capture differences in biomarker distributions, we proposed a novel method called QBP. Our method, that uses the quantiles of the continuous biomarker distributions, was compared with traditional statistical classification methods such as LR, PLR, PCLR, LDA and PLS-LDA, as well as more novel machine learning techniques such as kNN, RF, SVM and XGB. We studied the predictive performance of QBP compared to the alternative methods, but also other features, e.g. effect of sample size and number of selected biomarkers/components in the final model.  \\

In a simulation study, differences in means, variance and skewness between cases and controls were simulated for certain biomarkers. When cases and controls were drawn from the same distribution (dataset 1), it was demonstrated that QBP is unbiased (average $AUC=0.5$) just like all other methods. In the two datasets with biomarkers having only systematic shifts in the mean with a size of one times the standard deviation, LDA tends to be superior ($AUC=0.977$ in dataset 2 and $AUC=0.996$ in dataset 3). Compared to LDA, QBP has a worse predictive performance in terms of AUC ($AUC=0.854 \; (-12.6\%) \text{ and } AUC=0.948 \; (-5\%)$, for dataset 2 and 3 respectively). In contrast to the performance gap with LDA, PLR, PLS-LDA and SVM, QBP performs just slightly worse compared to RF and XGB. In case of normally distributed data with a shift in standard deviation (dataset 4), QBP is superior to all methods ($AUC=0.652$). Whereas RF and XGB seems to come relatively close ($AUC=0.629$ and $AUC=0.584$, respectively), all logistic regression and LDA based techniques and SVM.Linear fail to discriminate better than random ($AUC=0.5$). 

In order to create a mixture of skewed and not skewed biomarker distributions, both normal and log-normal biomarkers are simulated. When simulating a shift in skewness, while remaining the mean and variance constant (dataset 5), RF and XGB were superior ($AUC=0.963$ and $AUC=0.917$, respectively), followed by QBP ($AUC=0.860$). All other techniques show a very weak classification performance ($AUC<0.569$). Compared to dataset 2 and 3, it seems that changing the biomarker distribution from normally distributed biomarkers to log-normally distributed biomarkers -- while maintaining the shifts in mean parameter for some biomarkers (dataset 6a, 6b and 6c) -- just slightly changes the relative differences in performance between the techniques. In specific, QBP demonstrated an inferior performance ($AUC=0.688 \; (-14.4\%),\; AUC=0.785 \; (-9.3\%) \text{ and } AUC=0.724 \; (-13\%)$ for datasets 6a, 6b and 6c, respectively) relative to the best in class Lasso. Simultaneously, the gap between Lasso and the machine learning techniques RF and XGB has shrinked. In the datasets with only changes in the variances for some biomarkers and log-normal biomarker distributions (datasets 7a, 7b and 7c), the predictive performance of QBP ($AUC=0.652,\; AUC=0.751 \text{ and } AUC=0.674$ for datasets 7a, 7b and 7c, respectively) was better or equal compared to its successor RF. This conclusion is also in line with dataset 4, where the data was normally distributed. Note that the difference in performance between QBP and RF decreased with increasing sample size. In the datasets where biomarkers may change in means, in variance or in both (dataset 8a, 8b and 8c), QBP performed equal compared to RF and XGB and was superior in relation to the other methods in terms of prediction. Thus, in the most realistic setting -- where cases and controls do not just differ in mean -- QBP truly competes with XGB and RF and does substantially better than more classical methods.

The simulation study also showed for all methods that an increase in sample size tends to increase the predictive performance and decrease the standard deviation. In particular, a balanced increase of the number of cases and controls appeared to be most effective. A primary cause of this increased performance is the fact that the standard error of the quantiles decreases when increasing the sample size. For QBP, this directly results into more precise estimates for the quantiles and estimates of the exceedratios. As a consequence, the probability of falsely including biomarkers decreases. This sample size effect was mainly visible for QBP in the lower number of selected biomarkers and the increased specificity and accuracy of the biomarker selection for the balanced datasets with $n=400$ compared to $n=100$. For the PLR methods on the other hand, the specificity and accuracy decreased with increasing sample size, except for datasets 8a, 8b and 8c where the accuracy increased with increasing sample size. Whenever, a relative number of biomarkers is involved with different variances between cases and controls, QBP has a better sensitivity than traditional methods, although not always a better specificity when the number of cases and/or controls is low. This was observed in the balanced datasets with $n=100$.

Apart from the simulation study, two case studies were analyzed: a major depression disorder dataset and a trisomy dataset. Whereas the traditional methods barely detected any difference between cases and controls in the MDD dataset ($AUC \approx 0.5$), QBP reached an area under the curve of 0.680, which is more than $7.1 \%$ and $12.2\%$ higher than the two successors RF and XGB, respectively. This superior performance can mainly be ascribed to the fact that most relevant biomarkers in this dataset show differences in distributional characteristics other than just differences in means between cases and controls. When considering the predictive performance of the methods on the trisomy dataset using all biomarkers, it can be concluded that QBP ($AUC=0.908$) performs equally well as LR, PCLR, SVM.Linear and XGB, and significantly better than the other methods. A comparison of QBP and the FMF risk that is used by RIVM to predict trisomy was performed on a larger dataset with a lower number of biomarkers. It was shown that the classification performance of QBP in terms of the AUC is slightly better than the FMF risk ($AUC=0.9249$ and $AUC=0.9151$ for QBP and FMF risk, respectively). \\

In our simulation study, we only applied normal and log-normal distributions, but did not use other statistical distributions. However, QBP can easily be translated to other continuous statistical distributions, most likely without losing its strength in detecting tail differences. Moreover, note that in the implementation of PCLR, the principal components are selected in the natural order given by their explained variances. Although, an alternative method using a stepwise procedure of selecting principal components based on the conditional likelihood-ratio test is described to be superior \cite{aguilera2006using}, we do not expect the conclusions of this study to change in this case. We however used PLS-LDA as well, which creates sparse representation of the data before applying LDA. Finally, although we currently did not include interactions or other higher order terms, these could be easily constructed. \\

Additional research on the QBP should be conducted as the complete set of possible tunable parameters and corresponding settings have not been studied or explored in its full potential. This can be in terms of the number of percentiles and the corresponding proportions, where one could focus on its relation with the sample size. Note that the proportions should be selected with care, especially when dealing with small sample sizes, as this will result into less robust percentiles. Furthermore, it could be investigated whether the weights of biomarkers should be equal for all biomarkers or it should depend on a certain statistic. For example, biomarkers that vary in variation between cases and controls may receive larger weights that could be proportional to Levene's test of homogeneity. Thus it is not unlikely that the QBP can become even better in predicting cases and selecting relevant biomarkers. 

Another point of attention is the topic of collinearity, since it could easily inflate the disease scores of QBP. A simple precaution could be to reduce the biomarker weights of biomarker scores in case of confounding, however, more sophisticated measures could be developed. At the moment, QBP is limited to binary outcomes and continuous biomarkers. If one wants to include binary covariates such as gender or use multiple outcome levels this is not straightforward. For binary covariates, we could for example apply location-scale transformations. Especially in datasets that are too small for separate QBP analyses this might be useful. For discrete covariates -- which we treated as continuous covariates in the trisomy dataset -- a more sophisticated rule based on proportions could established to improve the performance of QBP. From a computational perspective, QBP algorithm is currently more computationally intensive than other classical statistical methods -- especially in comparison to (P)LR or LDA. Relative to machine learning techniques, QBP seems to perform comparable or better. Note that the processing times are particularly high for the techniques that require CV to select the optimal set of tunable parameters. This CV was performed such that each method received exactly the same split of the training data, and with that ensuring a fair comparison by giving each method the same information to fit a model. Besides that the computational efficiency could still be improved, a mathematical or theoretical underpinning of QBP is needed to demonstrate its capability. \\

Summarizing, QBP outperforms the observed traditional methods in discriminating cases from controls if the predictor variables show differences in variances between cases and controls. In case only systematic shifts in the mean of normally or log-normally distributed predictor variables are present, QBP is inferior to the traditional methods. For situations with mixtures of shifts in means, variances or other distributional differences, as expected in real life due to complex biological processes, QBP was superior to all methods in the MDD casestudy and was amongst the best performing methods in the simulation study -- together with RF and XGB. There are still numerous settings for which the performance of QBP should be assessed, but we demonstrated its potential on predicting diseases. Although QBP is currently applied on disease classification, it can be used in all fields involving binary classification with continuous covariates, such as economics, marketing, engineering and social sciences.

\section*{Acknowledgements}
The Foundation for Prenatal Screening in Northern Netherlands is gratefully acknowledged for providing the data for the Trisomy case study, enabling us to perform the analysis on a large set of routine clinical screening data.



\AtNextBibliography{\small \emergencystretch=15em}
\printbibliography

@article{tibshirani1996regression,
  title={Regression shrinkage and selection via the lasso},
  author={Tibshirani, Robert},
  journal={Journal of the Royal Statistical Society. Series B (Methodological)},
  pages={267--288},
  year={1996},
  publisher={JSTOR}
}

@article{everitt2011miscellaneous,
  title={Miscellaneous clustering methods},
  author={Everitt, Brian S and Landau, Sabine and Leese, Morven and Stahl, Daniel},
  journal={Cluster analysis},
  pages={215--255},
  year={2011},
  publisher={Wiley}
}

@article{coomans1982alternative,
  title={Alternative k-nearest neighbour rules in supervised pattern recognition: Part 1. k-Nearest neighbour classification by using alternative voting rules},
  author={Coomans, Danny and Massart, D{\'e}sir{\'e} Luc},
  journal={Analytica Chimica Acta},
  volume={136},
  pages={15--27},
  year={1982},
  publisher={Elsevier}
}

@book{world2001biomarkers,
  title={Biomarkers in risk assessment: Validity and validation},
  author={World Health Organization and others},
  volume={222},
  year={2001},
  publisher={WHO}
}

@Book{MASSpackage,
    title = {Modern Applied Statistics with S},
    author = {W. N. Venables and B. D. Ripley},
    publisher = {Springer},
    edition = {Fourth},
    address = {New York},
    year = {2002},
    note = {ISBN 0-387-95457-0},
    url = {http://www.stats.ox.ac.uk/pub/MASS4},
}

@article{ROCRpackage,
    entry = {article},
    title = {ROCR: visualizing classifier performance in R},
    author = {T. Sing and O. Sander and N. Beerenwinkel and T. Lengauer},
    year = {2005},
    journal = {Bioinformatics},
    volume = {21},
    number = {20},
    pages = {7881},
    url = {http://rocr.bioinf.mpi-sb.mpg.de},
}

@article{shiefa2013first,
  title={First trimester maternal serum screening using biochemical markers PAPP-A and free $\beta$-hCG for down syndrome, patau syndrome and edward syndrome},
  author={Shiefa, S and Amargandhi, M and Bhupendra, J and Moulali, S and Kristine, T},
  journal={Indian Journal of Clinical Biochemistry},
  volume={28},
  number={1},
  pages={3--12},
  year={2013},
  publisher={Springer}
}

@article{nicolaides2003screening,
  title={Screening for chromosomal defects},
  author={Nicolaides, KH},
  journal={Ultrasound in Obstetrics \& Gynecology},
  volume={21},
  number={4},
  pages={313--321},
  year={2003},
  publisher={Wiley Online Library}
}

@article{boulesteix2004pls,
  title={PLS dimension reduction for classification with microarray data},
  author={Boulesteix, Anne-Laure and others},
  journal={Statistical applications in genetics and molecular biology},
  volume={3},
  number={1},
  pages={1075},
  year={2004}
}

@article{de1993simpls,
  title={SIMPLS: an alternative approach to partial least squares regression},
  author={De Jong, Sijmen},
  journal={Chemometrics and intelligent laboratory systems},
  volume={18},
  number={3},
  pages={251--263},
  year={1993},
  publisher={Elsevier}
}

@article{vera2011discrimination,
  title={Discrimination and sensory description of beers through data fusion},
  author={Vera, Luciano and Ace{\~n}a, L and Guasch, J and Boqu{\'e}, R and Mestres, M and Busto, O},
  journal={Talanta},
  volume={87},
  pages={136--142},
  year={2011},
  publisher={Elsevier}
}

@article{nguyen2002tumor,
  title={Tumor classification by partial least squares using microarray gene expression data},
  author={Nguyen, Danh V and Rocke, David M},
  journal={Bioinformatics},
  volume={18},
  number={1},
  pages={39--50},
  year={2002},
  publisher={Oxford Univ Press}
}

@article{marigheto1998comparison,
  title={A comparison of mid-infrared and Raman spectroscopies for the authentication of edible oils},
  author={Marigheto, NA and Kemsley, EK and Defernez, M and Wilson, RH},
  journal={Journal of the American oil chemists’ society},
  volume={75},
  number={8},
  pages={987--992},
  year={1998},
  publisher={Springer}
}

@article{just2014improving,
  title={Improving tumour heterogeneity MRI assessment with histograms},
  author={Just, Nathalie},
  journal={British journal of cancer},
  volume={111},
  number={12},
  pages={2205--2213},
  year={2014},
  publisher={Nature Publishing Group}
}

@article{wold1985partial,
  title={Partial least squares},
  author={Wold, Herman},
  journal={Encyclopedia of statistical sciences},
  year={1985},
  publisher={Wiley Online Library}
}

@article{hoerl1970ridge,
  title={Ridge regression: Biased estimation for nonorthogonal problems},
  author={Hoerl, Arthur E and Kennard, Robert W},
  journal={Technometrics},
  volume={12},
  number={1},
  pages={55--67},
  year={1970},
  publisher={Taylor \& Francis Group}
}

@article{westerhuis2008assessment,
  title={Assessment of PLSDA cross validation},
  author={Westerhuis, Johan A and Hoefsloot, Huub CJ and Smit, Suzanne and Vis, Daniel J and Smilde, Age K and van Velzen, Ewoud JJ and van Duijnhoven, John PM and van Dorsten, Ferdi A},
  journal={Metabolomics},
  volume={4},
  number={1},
  pages={81--89},
  year={2008},
  publisher={Springer}
}

@article{filzmoser2009repeated,
  title={Repeated double cross validation},
  author={Filzmoser, Peter and Liebmann, Bettina and Varmuza, Kurt},
  journal={Journal of Chemometrics},
  volume={23},
  number={4},
  pages={160--171},
  year={2009},
  publisher={Wiley Online Library}
}

@article{pepe2008pivotal,
  title={Pivotal evaluation of the accuracy of a biomarker used for classification or prediction: standards for study design},
  author={Pepe, Margaret S and Feng, Ziding and Janes, Holly and Bossuyt, Patrick M and Potter, John D},
  journal={Journal of the National Cancer Institute},
  volume={100},
  number={20},
  pages={1432--1438},
  year={2008},
  publisher={Oxford University Press}
}

@article{aguilera2006using,
  title={Using principal components for estimating logistic regression with high-dimensional multicollinear data},
  author={Aguilera, Ana M and Escabias, Manuel and Valderrama, Mariano J},
  journal={Computational Statistics \& Data Analysis},
  volume={50},
  number={8},
  pages={1905--1924},
  year={2006},
  publisher={Elsevier}
}

@article{hsu2014biomarker,
  title={Biomarker selection for medical diagnosis using the partial area under the ROC curve},
  author={Hsu, Man-Jen and Chang, Yuan-Chin Ivan and Hsueh, Huey-Miin},
  journal={BMC research notes},
  volume={7},
  number={1},
  pages={1},
  year={2014},
  publisher={BioMed Central}
}

@article{sobocki2006cost,
  title={Cost of depression in Europe.},
  author={Sobocki, Patrik and J{\"o}nsson, Bengt and Angst, Jules and Rehnberg, Clas},
  journal={The journal of mental health policy and economics},
  volume={9},
  number={2},
  pages={87--98},
  year={2006}
}

@article{bromet2011cross,
  title={Cross-national epidemiology of DSM-IV major depressive episode},
  author={Bromet, Evelyn and Andrade, Laura Helena and Hwang, Irving and Sampson, Nancy A and Alonso, Jordi and De Girolamo, Giovanni and De Graaf, Ron and Demyttenaere, Koen and Hu, Chiyi and Iwata, Noboru and others},
  journal={BMC medicine},
  volume={9},
  number={1},
  pages={1},
  year={2011},
  publisher={BioMed Central}
}

@article{calfee2011use,
  title={Use of risk reclassification with multiple biomarkers improves mortality prediction in acute lung injury},
  author={Calfee, Carolyn S and Ware, Lorraine B and Glidden, David V and Eisner, Mark D and Parsons, Polly E and Thompson, B Taylor and Matthay, Michael A and others},
  journal={Critical care medicine},
  volume={39},
  number={4},
  pages={711},
  year={2011},
  publisher={NIH Public Access}
}

@article{schielen2006multi,
  title={Multi-centre first-trimester screening for Down syndrome in the Netherlands in routine clinical practice},
  author={Schielen, PCJI and van Leeuwen-Spruijt, M and Belmouden, I and Elvers, LH and Jonker, M and Loeber, JG},
  journal={Prenatal diagnosis},
  volume={26},
  number={8},
  pages={711--718},
  year={2006},
  publisher={Wiley Online Library}
}

@article{jentsch2015biomarker,
  title={Biomarker approaches in major depressive disorder evaluated in the context of current hypotheses},
  author={Jentsch, Mike C and Van Buel, Erin M and Bosker, Fokko J and Gladkevich, Anatoliy V and Klein, Hans C and Voshaar, Richard C Oude and Ruh{\'e}, Henricus G and Eisel, Uli LM and Schoevers, Robert A},
  journal={Biomarkers},
  volume={9},
  number={3},
  pages={277--297},
  year={2015},
  publisher={Future Medicine}
}

@article{hosmer2000introduction,
  title={Introduction to the logistic regression model},
  author={Hosmer, David W and Lemeshow, Stanley},
  journal={Applied Logistic Regression, Second Edition},
  pages={1--30},
  year={2000},
  publisher={Wiley Online Library}
}

@article{bradley1997use,
  title={The use of the area under the ROC curve in the evaluation of machine learning algorithms},
  author={Bradley, Andrew P},
  journal={Pattern recognition},
  volume={30},
  number={7},
  pages={1145--1159},
  year={1997},
  publisher={Elsevier}
}

@article{smit2007assessing,
  title={Assessing the statistical validity of proteomics based biomarkers},
  author={Smit, Suzanne and van Breemen, Mari{\"e}lle J and Hoefsloot, Huub CJ and Smilde, Age K and Aerts, Johannes MFG and De Koster, Chris G},
  journal={Analytica Chimica Acta},
  volume={592},
  number={2},
  pages={210--217},
  year={2007},
  publisher={Elsevier}
}

@article{halaris2013inflammation,
  title={Inflammation, heart disease, and depression},
  author={Halaris, Angelos},
  journal={Current psychiatry reports},
  volume={15},
  number={10},
  pages={1--9},
  year={2013},
  publisher={Springer}
}

@article{ma2008penalized,
  title={Penalized feature selection and classification in bioinformatics},
  author={Ma, Shuangge and Huang, Jian},
  journal={Briefings in bioinformatics},
  volume={9},
  number={5},
  pages={392--403},
  year={2008},
  publisher={Oxford Univ Press}
}

@article{zou2005regularization,
  title={Regularization and variable selection via the elastic net},
  author={Zou, Hui and Hastie, Trevor},
  journal={Journal of the Royal Statistical Society: Series B (Statistical Methodology)},
  volume={67},
  number={2},
  pages={301--320},
  year={2005},
  publisher={Wiley Online Library}
}

@article{steyerberg2010assessing,
  title={Assessing the performance of prediction models: a framework for some traditional and novel measures},
  author={Steyerberg, Ewout W and Vickers, Andrew J and Cook, Nancy R and Gerds, Thomas and Gonen, Mithat and Obuchowski, Nancy and Pencina, Michael J and Kattan, Michael W},
  journal={Epidemiology (Cambridge, Mass.)},
  volume={21},
  number={1},
  pages={128},
  year={2010},
  publisher={NIH Public Access}
}

@book{zhou2012ensemble,
  title={Ensemble methods: foundations and algorithms},
  author={Zhou, Zhi-Hua},
  year={2012},
  publisher={CRC press}
}

@inproceedings{chen2016xgboost,
  title={Xgboost: A scalable tree boosting system},
  author={Chen, Tianqi and Guestrin, Carlos},
  booktitle={Proceedings of the 22nd acm sigkdd international conference on knowledge discovery and data mining},
  pages={785--794},
  year={2016}
}

@book{friedman2001elements,
  title={The elements of statistical learning},
  author={Friedman, Jerome and Hastie, Trevor and Tibshirani, Robert},
  volume={1},
  number={10},
  year={2001},
  publisher={Springer series in statistics New York}
}

@article{cortes1995support,
  title={Support-vector networks},
  author={Cortes, Corinna and Vapnik, Vladimir},
  journal={Machine learning},
  volume={20},
  number={3},
  pages={273--297},
  year={1995},
  publisher={Springer}
}

@inproceedings{boser1992training,
  title={A training algorithm for optimal margin classifiers},
  author={Boser, Bernhard E and Guyon, Isabelle M and Vapnik, Vladimir N},
  booktitle={Proceedings of the fifth annual workshop on Computational learning theory},
  pages={144--152},
  year={1992}
}

@INPROCEEDINGS{Platt99probabilisticoutputs,
    author = {John C. Platt},
    title = {Probabilistic Outputs for Support Vector Machines and Comparisons to Regularized Likelihood Methods},
    booktitle = {ADVANCES IN LARGE MARGIN CLASSIFIERS},
    year = {1999},
    pages = {61--74},
    publisher = {MIT Press}
}

@techreport{friedman1999stochastic,
  title={Stochastic gradient boosting. Department of Statistics},
  author={Friedman, JH},
  year={1999},
  institution={Stanford University, Technical Report, San Francisco, CA}
}

@article{glmnet,
    title = {Regularization Paths for Generalized Linear Models via
      Coordinate Descent},
    author = {Jerome Friedman and Trevor Hastie and Robert Tibshirani},
    journal = {Journal of Statistical Software},
    year = {2010},
    volume = {33},
    number = {1},
    pages = {1--22},
    url = {http://www.jstatsoft.org/v33/i01/},
  }


\end{document}